\documentclass[review]{elsarticle}

\usepackage{hyperref}
\usepackage[T1]{fontenc}
\usepackage[utf8]{inputenc}
\usepackage{graphicx}
\usepackage{amsmath}
\usepackage{algorithm, algpseudocode}
\usepackage{subcaption}
\usepackage{adjustbox}

\journal{Applied Soft Computing}

\begin{document}

\begin{frontmatter}

\title{A Reinforcement Learning Based Encoder-Decoder Framework for Learning Stock Trading Rules}


\author[mymainaddress,mysecondaryaddress]{Mehran Taghian}
\ead{mehrantaghian@aut.ac.ir}
\author[mymainaddress,mysecondaryaddress]{Ahmad Asadi}
\ead{ahmad.asadi@aut.ac.ir}
\author[mymainaddress,mysecondaryaddress]{Reza Safabakhsh\corref{mycorrespondingauthor}}
\cortext[mycorrespondingauthor]{Corresponding author}
\ead{safa@aut.ac.ir}

\begin{abstract}
    	A wide variety of deep reinforcement learning (DRL) models have recently been proposed to learn profitable investment strategies. The rules learned by these models outperform the previous strategies specially in high frequency trading environments. However, it is shown that the quality of the extracted features from a long-term sequence of raw prices of the instruments greatly affects the performance of the trading rules learned by these models. Employing a neural encoder-decoder structure to extract informative features from complex input time-series has proved very effective in other popular tasks like neural machine translation and video captioning in which the models face a similar problem. The encoder-decoder framework extracts highly informative features from a long sequence of prices along with learning how to generate outputs based on the extracted features. In this paper, a novel end-to-end model based on the neural encoder-decoder framework combined with DRL is proposed to learn single instrument trading strategies from a long sequence of raw prices of the instrument. The proposed model consists of an encoder which is a neural structure responsible for learning informative features from the input sequence, and a decoder which is a DRL model responsible for learning profitable strategies based on the features extracted by the encoder. The parameters of the encoder and the decoder structures are learned jointly, which enables the encoder to extract features fitted to the task of the decoder DRL. In addition, the effects of different structures for the encoder and various forms of the input sequences on the performance of the learned strategies are investigated. Experimental results showed that the proposed model outperforms other state-of-the-art models in highly dynamic environments. 
\end{abstract}


\end{frontmatter}

    \section{Introduction}
\label{sec:intro}
Forming profitable trading strategies fitted on either a single financial instrument or a set of instruments in a specific market, based on a vast historical data is a critical problem for investors. Since the introduction of algorithmic trading \cite{chan2009quantitative} and monitoring the trading process by computers, especially at high frequency \cite{gomber2015high}, there has been a widespread interest in designing a powerful model to learn profitable investment strategies. 

In recent years, machine learning (ML) models and deep neural networks (DNNs) have been widely used for learning profitable investment strategies in both single asset trading and portfolio management problems \cite{zhang2020deep}. Among the techniques used for learning asset-specific trading rules, genetic programming (GP) and deep reinforcement learning (DRL) methods have been more interesting for the research community\cite{taghian2020learning}.
 
Genetic programming was widely used to learn technical trading rules for different indices like S\&P500 index (\cite{allen1999using}), to learn appropriate trading rules to benefit from short-term price fluctuations (\cite{potvin2004generating}), to learn noise-tolerant rules based on a large number of technical indicators (\cite{chien2010mining}), and to learn the trading rules based on popular technical indicators like MACD (\cite{mallick2008empirical}). 

Since genetic programming and its modifications are not able to evolve after task execution, reinforcement learning techniques have been widely used to combine with evolutionary algorithms to cover this weakness. \cite{chen2007genetic} combined the SARSA algorithm with genetic programming to enable the model to change programs during task execution. \cite{yang2010gnp} modified the GNP-SARSA model proposed by \cite{chen2007genetic} with augmenting new nodes called subroutines to make an appropriate trade-off between the efficiency and compactness of the model.

Considering the great performance of deep reinforcement learning (DRL) models (deep neural networks trained with the reinforcement learning techniques) in forming investment strategies, the proposed methods for portfolio management are mainly based on DNN structures and DRL techniques. \cite{ganesh2018deep} proposed a very-long short-term memory (VLSTM) network, to deal with extremely long sequences in financial markets and explored the importance of VLSTM in the context of high frequency trading. \cite{arevalo2016high} proposed a DNN structure to forecast the next one-minute average price of an instrument given its current time and $n$-lagged one-minute pseudo-returns to build a trading strategy that buys (sells) when the next predicted average price is above (below) the last closing price. \cite{dixon2019deep} proposed a DNN model to learn the spatio-temporal model of the input and developed a classification rule to predict short-term futures market prices using order book depth. 

\cite{rundo2019deep} proposed a DRL technique for forecasting the short-term trend in the currency FOREX (FOReign EXchange) market to maximize the return on investment in an HFT algorithm. Jiang et al. \cite{jiang2017cryptocurrency} proposed a financial model-free reinforcement learning framework to provide a deep learning solution to the portfolio management problem. For single stocks trading, Wang et al. \cite{wang2017deep} employed deep Q-learning to build an end-to-end deep Q-trading system for learning trading strategies.

\cite{taghian2020learning} studied the DRL performance in learning single asset-specific trading rules and concluded that: 1) the quality of extracted features from the input can greatly affect the performance of the learned strategy by DRL models, and 2) proposing a good feature extractor from a long-term historical price data sequence would obviously improve the profitability of the resulting trading strategy. Considering the results from \cite{taghian2020learning}, proposing a model to learn good features from a long-term price sequence would effectively contribute to improve the performance of DRL models.

Considering the results reported by \cite{taghian2020learning}, we combined the encoder-decoder framework with the DRL techniques and proposed an end-to-end model to learn informative features from a long-term price sequence of a specific financial instrument and learn a profitable trading strategy. The encoder-decoder framework is one of the state-of-the-art neural structures applied in tasks requiring extracting complex feature representation, specially in cases that the input is presented in the form of a long-term time-series\cite{asadi2020encoder}. 

%

In this paper, we first develop a DRL agent based on a deep Q-learning algorithm to generate trading signals given a sequence of OHLC prices of each instrument. Then, we design and implement an encoder-decoder based model to improve the agent's feature extraction performance. In addition, we examine the performance of different DNN structures for the encoder module. The time-series of candlesticks and raw OHLC input types are evaluated, and the performance of models is tested using various stocks with different behavior. Furthermore, the influence of window size on the agent's performance for the windowed input type has been studied. Experimental results showed that our model outperforms the state-of-the-art methods.

In the next section of this paper, we briefly review the related work of learning financial asset-specific trading strategies and discuss the advantages and disadvantages of different categories of the proposed methods. Section 3, discusses the model proposed in this paper. The model consists of an encoder part for feature extraction and a decoder part for decision making. The details of the architecture of both of these parts are discussed in Section 3. Section 4 provides the experimental results and the conclusions are provided in Section 5.

\section{Related Work}
\label{sec:related}



many researchers proposed methods based on reinforcement learning for determining trading strategies. \cite{moody1998performance} first applied RL in portfolio management, and proposed a method based on recurrent reinforcement learning for financial transactions. \cite{jangmin2006adaptive} proposed a method based on the RL framework which incorporates stock selection and asset management. 
\cite{suchaimanacharoen2020empowered} incorporate time-series prediction with the decision making power of RL. They first predict the future prices using a CNN, and then feed the output to a policy gradient model along with historical data to empower trading decisions. 

\cite{mnih2015human} first incorporate deep neural networks to learn policy directly from high dimensional sensory inputs. The proposed method, termed deep Q-learning, successfully played seven different Atari games, three of which could outperform the human level. 
Considering the successful performance of deep reinforcement learning in playing Atari games, researchers have carried out many kinds of research to apply DRL methods to the stock market environment.


\cite{luo2019novel} applied two different CNN based function approximators with an actor-critic RL algorithm called “Deep Deterministic Policy Gradient” (DDPG) to find the optimal policy. The proposed DDPG has two different convolutional neural network (CNN) function approximators. The input state to the model is 18 different technical indicators converted to multiple channels of 1D images fed to the CNN model.

\cite{wang2019alphastock} proposed a novel RL based investment strategy consisting of three phases: 1) extracting asset representation from multiple time-series using a Long Short Term Memory with a History Attention(LSTM-HA) network, 2) modeling the interrelationships among assets as well as the asset price rising prior using a Cross-Asset Attention Network(CAAN), and 3) generating portfolio and giving the investment proportion of each asset according to the output winner scores of the attention network. The three components are optimized end-to-end using a Sharpe ratio oriented RL.

\cite{xiong2018practical} explored the training power of the Deep Deterministic Policy Gradient to learn stock trading strategy. \cite{chakole2020q} proposed a method using the Q-learning algorithm to find the optimal dynamic trading strategy. They introduced two models varied in their representation of the environment, the first of which represents environment states using a finite set of clusters, the second of which used the candlesticks themselves as the states of the environment.

\cite{theate2020application} presented a solution to the algorithmic trading problem of generating the trading strategy for single stock based on the DQN algorithm with a Sharpe ratio oriented manner. \cite{brim2020deep} used co-integrated stock market prices and incorporated DQN to generate pairs trading strategy.

Following the methods attempting to apply RL and DRL techniques to generate stock market trading strategies and portfolio management, some methods tried to improve various parts of the agent and environment to improve the RL agent's performance. 

\cite{calabuig2020dreaming} used McShane-Whitney extension of the Lipschitz function (\cite{jimenez2020mcshane}) to forecast the reward function based on its previous values. Besides, to support the extension of the reward function, all the previous states, along with some artificially generated states, called dream states, were combined to enrich learning.  

\cite{park2020intelligent} proposed an approach based on deep Q-learning for deriving a multi-asset portfolio trading strategy. Instead of using a discrete action-space, which might lead to infeasible actions, they introduced a mapping function to map unreasonable and infeasible actions to similar valuable actions. Therefore, the trading strategy would be more reasonable in the practical action space. Besides, the dimensionality problem of the action space is taken care of by using deep Q-learning.

Having considered the temporal essence of stock market data, some of the researches have combined the temporal feature extraction power of recurrent neural networks with DRL's decision-making ability. \cite{wu2020adaptive} applied the Gated Recurrent Unit(GRU) to exploit informative features from raw financial data along with technical indicators to represent stock market conditions more robustly. Then they designed a risk-adjusted reward function using the Sortino ratio proposed by \cite{rollinger2013sortino}. Based on the state, action, and reward functions designed, they proposed Deep Q-Learning and Deep Deterministic Policy Gradient for quantitative stock trading.

\cite{weng2020portfolio} applied DRL for portfolio management and in order to distinguish the critical time when the price changes, they proposed using a three-dimensional attention gating network that gave higher weights on rising moments and assets. Moreover, they applied the XGBoost method to quantify the importance of features and output the three most relevant features from historical data to the model: close price, high price, and low price.

\cite{pendharkar2018trading} compared different RL agents for trading financial indices in a personal retirement portfolio. The comparison included on-policy SARSA($\lambda$) and off-policy Q($\lambda$) with discrete state and discrete action settings that maximize either total return or differential Sharpe ratios, on-policy temporal difference learning, and TD($\lambda$) with discrete state and continuous action settings. They showed that an adaptive continuous action agent has the best performance in predicting next period portfolio allocations.

\cite{almahdi2019constrained} incorporated particle swarm optimization algorithm to optimize portfolio. Besides, they developed a recurrent reinforcement learning based method for portfolio allocation and trading.

In our former work (\cite{taghian2020learning}), we studied the performance of strategies based on the candlestick patterns, SARSA($\lambda$) algorithm, and deep Q-learning and concluded that methods based o deep reinforcement learning could generate more adaptive trading strategies specific to each asset. 



The encoder-decoder framework, is one of the most popular neural structures which is used to solve complex problems in an end-to-end menner (\cite{asadi2020encoder}).This architecture was first proposed in neural machine translation by \cite{cho2014learning}. It consists of two sequential parts: 1) An encoder part which is responsible to lean a good feature extraction from the input and 2) a decoder which is responsible for generating appropriate output based on the extracted feature vector.

Since employing the encoder-decoder framework in neural machine translation has significantly improved the performance of the models, it is widely used in other complex tasks like image captioning (see \cite{lu2017knowing} and \cite{xu2015show} for more details), and video captioning (see \cite{donahue2015long} for more details).

In this work, we dedicate focus on the following issues :
\begin{enumerate}
	\item Proposing a model based on the encoder-decoder architecture where the decoder is a DRL agent trained to generate a trading strategy based on the representation of the market produced by the encoder model. The encoder is a neural network structure responsible for exploiting features from the raw input time-series data and generating feature vector.
	\item Showing the importance of extracting time dependencies of the input prices, and proposing different encoder structures are proposed to improve the quality of time dependencies extraction.
	\item Investigating the impact of window-size on exploiting important features and generating a proper representation of the market.
\end{enumerate}

The rest of this paper is structured as follows: first, we introduce the details of the proposed method, deep Q-learning method, model architecture, and DNN models used as encoder. Then, the performance of different encoder models are evaluated using various methods explained in detail in Section 4. Next, the results of experiments are analyzed.

\section{Proposed Method}
\label{sec:method}

\subsection{Formulation}
\begin{figure}
	\centering
	\includegraphics[scale=0.2]{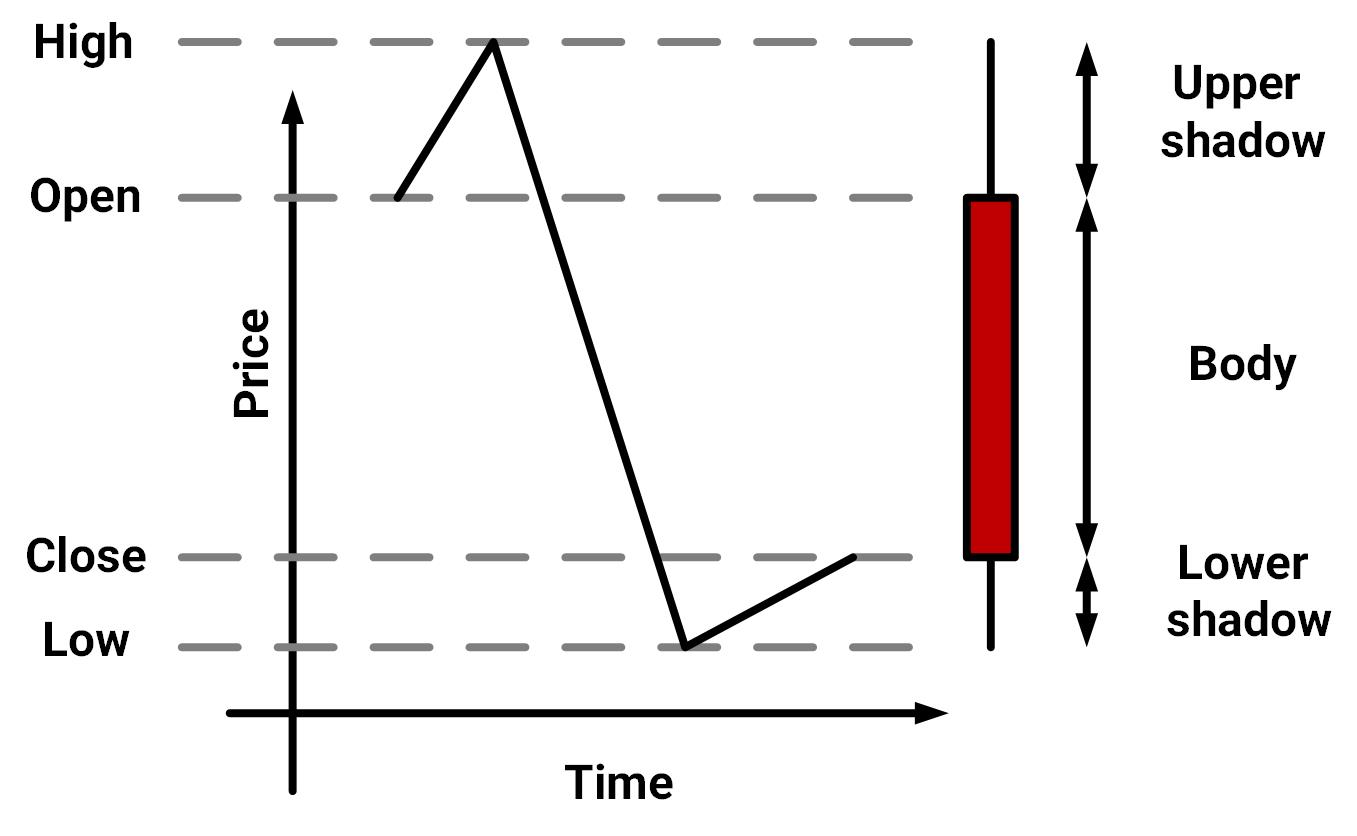}
	\caption{A candlestick representing the price behavior of an asset during a specific time window \cite{taghian2020learning}}
	\label{fig:candle}
\end{figure}

In financial markets, a candlestick is used to represent the price fluctuations in a short time period, originating from Japanese rice traders and merchants who used candlesticks to track the market prices\cite{northcott2009complete}. A candlestick consists of 4 price elements, namely high (the highest stock price during a period - e.g., a day), low (the lowest price), open (the stock price at the beginning of the period), and close(the stock price at the end of the period), abbreviated to OHLC. A candlestick's color can be either green/white, representing a bullish candle(open price is lower than close price), or red/black representing a bearish candle(close price is lower than open price). \ref{fig:candle} shows a sample candlestick. \eqref{eq:candle} shows a vector representing a candlestick. This vector consists of the Open, High, Low, and Close prices. 

\begin{equation}
c_t = (p_{open}, p_{high}, p_{low}, p_{close})
\label{eq:candle}
\end{equation}

A candlestick chart is used to demonstrate the behavior of the asset price. According to this concept, patterns in this chart show the buyers' and sellers' behavior and their influence on the market. Thus, these patterns can be used to analyze the price fluctuations and use the analysis to devise trade strategies on a financial asset. 

\subsection{Model architecture}

The proposed model is based on the encoder-decoder framework which consists of following modules:
\begin{enumerate}
	\item Encoder:\\
	Encoder is the first neural structure that takes the input of the model and learns a good mapping from the input space to the feature space which minimizes the decoders' loss function.
	\item Decoder:\\
	Decoder is the second neural structure in the encoder-decoder framework, that takes the features extracted by the encoder for each input record and generates the appropriate output based on the input feature vector. The gradients of the decoder are back propagated to the encoder and train its weights along with the weights of the decoder during the training phase.
\end{enumerate} 

In the proposed model, the decoder part is a target or policy network used in the deep Q-Learning based model proposed by \cite{taghian2020learning} to learn trading strategies. The encoder part is a deep neural network applied to extract deep features from the candlestick chart representations. These features are categorized into the following two groups:

\begin{enumerate}
	\item Features directly learned from candlestick representations or raw OHLC data.
	\item Features representing the temporal relationships among a sequence of candlesticks inside a time window.
\end{enumerate}

\begin{figure*}[htb]
	\centering
	\includegraphics[width=0.7\textwidth]{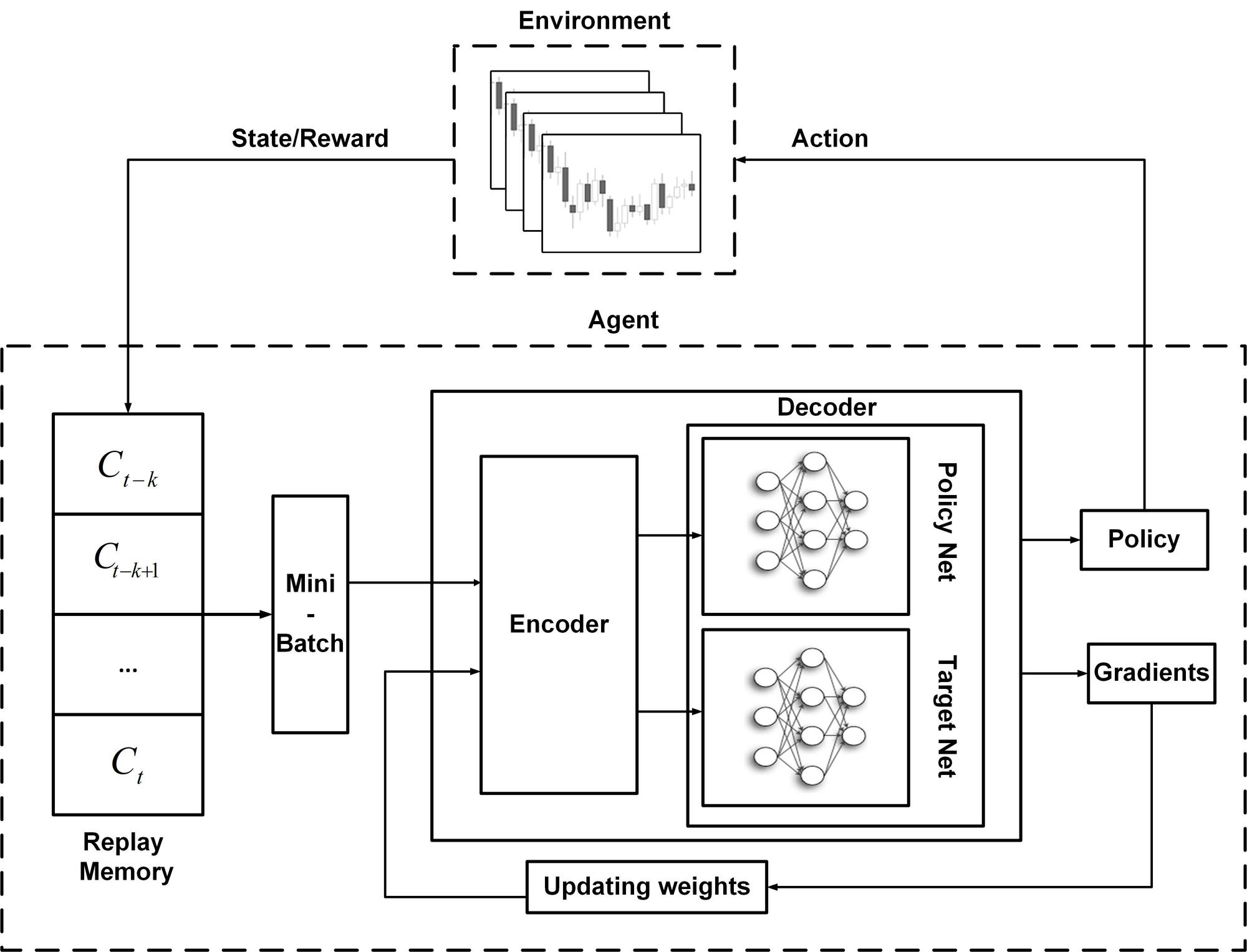}
	\caption{In this architecture, the state is given by the environment at each time step, and the agent takes action according to the state and receives the reward and next state. The quadruples $(CurrentState, Action, Reward, NextState)$ are saved in the replay memory. For the optimization part in each iteration, a batch of quadruples $(CurrentState, Action, Reward, NextState)$ is selected for the training after the step mentioned earlier. The encoder part's output is common between policy and target networks and is optimized in each iteration. The replay memory has a specific capacity, and after it is filled, a random quadruple is substituted with a new quadruple}
	\label{fig:arch}
\end{figure*}

For each category, NN models exist to efficiently analyze and extract those features according to the Policy Network's performance in trading. The encoder part extracts features from the input and provides a state vector (feature vector) for the decoder module, a deep Q-Learning agent that uses this state vector as the state of the environment to produce trading signals. Based on the rewards given to the DRL agent, the model is optimized towards producing higher profits. This optimization is done in an end-to-end form, back-propagating error from the decoder part to the encoder module. As a result, the encoder extracts features based on the trading performance of the DRL. The model architecture is shown in \ref{fig:arch}.

\subsection{The decoder module}
The decoder module is the trading agent of our model, and learns to produce trading signals. This module's architecture is based on the deep Q-Learning algorithm proposed by \cite{mnih2015human} to play Atari games. The module's input is a state vector, containing the features of the market at time-step t. These features can be either vanilla input of OHLC prices or the feature vector produced by a feature extractor model. We show the vector space of the input candles with $C = \{c_1, c_2, \dots, c_T\}$, where $T$ is the final time-step, and $c_i$ is the candle representation of different time-intervals (here daily). The feature extractor module $\phi$ gets the vector space of candles as input, and generates the vector space of states $S = \{s_1, s_2, \dots, s_T\}$. 

\begin{equation}
	\phi(C) = S
	\label{eq:feature-extractor}
\end{equation}

\eqref{eq:feature-extractor} represents $\phi$, the feature extraction function. This function can be either an identity function (state space is the candlesticks themselves) or a neural network that extracts deep features from the input vector space of candlesticks and produces a vector space of features. Each vector $s_i$ denotes the state of the environment at time-step t. This state is given to the DQN agent to use in taking action. The action space of the DQN agent is $A = \{'buy', 'sell', 'noop'\}$, which are the signals of the trading strategy at each time-step. After taking action, the agent would be given a reward based on the signal produced. The rewards of buying and selling are a bit different. \eqref{eq:reward} demonstrates the reward function used by the environment. The $ownShare$ parameter is used alongside action \textit{'noop'} showing that whether the money has already been invested on the market or not.

\begin{equation}
	R_t = \begin{cases}
		\begin{split}
			((1 - TC)^2 \times &\frac{P_2}{P_1} - 1) \times 100 \\ 
			&\text{if action = 'buy' or} \\
				&\text{(action = 'noop' and ownShare = True)}\\
		\end{split}\\\\
		\begin{split}
			((1 - TC)^2 \times &\frac{P_1}{P_2} - 1) \times 100 \\ 
			&\text{if action = 'sell' or} \\
			&\text{(action = 'noop' and ownShare = False)}\\
		\end{split}
	\end{cases}
	\label{eq:reward}
\end{equation}

Reinforcement learning is a framework used to learn a sequence of decision tasks. In general, the RL agent interacts with the environment, observes the state, takes action according to the policy and the observed state, and gets a reward. In a sequence of decisions made by the RL agent, the agent learns a policy $\pi$ regarding the actions taken and rewards earned at each episode. Afterward, the agent should optimize its policy to maximize cumulative reward after each episode. For this purpose, we used deep Q-learning, a critic based reinforcement learning algorithm, which uses action-value function Q(S, A) denoting the expected cumulative reward in state S when action A is taken. More formally, we use a multi-layered Perceptron to approximate the optimal action-value function, which is demonstrated in \eqref{eq:optimal-q}.

\begin{equation}
	Q^*(s, a) = \max_{\pi} E[r_t + \gamma r_{t+1} + \gamma^2 r_{t+2} + ... | s_t = s, a_t = a, \pi]
	\label{eq:optimal-q}
\end{equation}

where $\gamma$ is the discounting factor, $r_t$ is the reward at time-step t, $\pi$ is the behavior policy learned, $s$ is the observed state, and $a$ is the action taken.
The optimal action-value function obeys the Bellman equation:

\begin{equation}
	Q^*(s,a) = E_{s'} [r + \gamma \max_{a'} Q^*(s', a')|s,a]
	\label{eq:bellman}
\end{equation}

In order to reduce the mean squared error in the Bellman equation, we use two sets of parameters (Neural Networks). The target values are approximated using the target network weights (from previous iterations) shown with $\theta_i^-$ at iteration i. The policy network, which is being trained in each iteration to adjust its parameters to reduce the mean squared error, is used to approximate the Q function using parameters $\theta_i$ at iteration i. Thus, we have a sequence of loss functions $L_i(\theta_i)$ that changes at each iteration.

\begin{equation}
\begin{split}
	\label{eq:loss}
	L_i(\theta_i) &= E_{s, a, r}[(E_{s'}[y|s, a] - Q(s, a; \theta_i))^2] \\
	&= E_{s, a, r, s'}[(y - Q(s, a; \theta_i))^2] + E_{s, a, r}[V_{s'}[y]]
\end{split}
\end{equation}

In order to further stabilize the deep Q-learning algorithm, we use the Huber loss proposed by \cite{huber1992robust} instead of the Mean Squared Error, which pays attention to large and small errors equally. 

\begin{equation}
	Huber(e)= \begin{cases}
		\frac{1}{2}e^2 \quad\text{for } |e| \leq 1\\
		(|e| - \frac{1}{2}), \quad\text{otherwise} 
	\end{cases}
	\label{eq:huber}
\end{equation}

Furthermore, our agent stores the last $c$ experiences in the replay memory. 
The agent's experience vector $e_t = (s_t, a_t, r_t, s_{t+1})$ is saved in the Experience Replay Memory $D_t=\{e_1, e_2, \dots, e_c\}$ (where $c$ is the length of the Replay Memory) and used as a batch to optimize the policy network. When the model wants to update, it samples a batch of experiences uniformly at random from D. The steps of the deep Q-Learning algorithm used in our work are represented in Algorithm \ref{alg:q-learning}.

\begin{minipage}[htb]{\linewidth}
	\begin{algorithm}[H]
		\caption{Deep Q-Learning Algorithm used for training the agent}\label{alg:q-learning}
		\begin{algorithmic}[1]
			\State Initialize replay memory D to capacity N
			\State Initialize action-value function Q with random weights $\theta$
			\State Initialize target action-value function $\hat{Q}$ with weights $\theta^- = \theta$
			
			\For{episode from 1 to M}
			\State Initialize sequence $s_1$ and preprocessed sequence $\phi_1 = \phi(s1)$
			\For{t from 1 to T}
			\State With probability $\epsilon$ select a random action $a_t$
			\State Otherwise select $a_t = argmax_a Q(\phi(s_t), a; \theta)$
			\State Execute action $a_t$ and observe reward $r_t$ and state $s_{t+1}$
			\State Set $s_{t+1} = s_t, a_t$ and preprocess $\phi_{t+1} = \phi(s_{t+1})$
			\State Store transition ($\phi_t, a_t, r_t, \phi_{t+1}$)
			\State Sample random mini-batch of transitions ($\phi_j, a_j, r_j, \phi_{j+1}$) from D
			
			\State \begin{align*}
			Set \, y_j =
			\begin{cases}
			r_j \quad\text{if episode terminates at step j + 1}&\\
			r_j + \gamma \max_{a'} \hat{Q}(\phi_{j+1}, a'; \theta^-) \quad\text{otherwise}& \\
			\end{cases}
			\end{align*}
			
			\State Perform a gradient descent step on $(y_j - Q(\phi_j, a_j; \theta))^2$
			with respect to the network parameters $\theta$
			\State Every C steps reset $\hat{Q} = Q$
			\EndFor
			\EndFor
		\end{algorithmic}
	\end{algorithm}
\end{minipage}

\subsection{The encoder part}
So far, we have discussed the input data representation, different parts of the trading agent, and the deep Q-Learning algorithm, used by the agent to optimize its policy to learn devise profitable strategies. However, the essential part of each RL algorithm is the representation of the environment. As mentioned before, the environment tells the RL agent in which state it currently is, based on which the agent would take actions and receive rewards from the environment. A proper state representation can significantly improve the performance of the RL agent. Therefore, in this section, we want to concentrate on our model's feature extraction and state representation -- the Encoder module. 

We introduced the $\phi$ function, which given the input candlesticks, extracts features and outputs the state space, which is then fed to the decoder module -- the DQN model. The $\phi$ function can be either an identity function or a deep neural network. We do not have any feature extraction in the first case, and candlesticks are directly fed to the DQN model. The DNNs we use as the feature extractor are Multi-Layered Perceptron (MLP), Gated Recurrent Unit (GRU)\cite{cho2014learning}, 1-dimension Convolution in the direction of time (CNN)\cite{lecun1998gradient}, and GRU with 1-dimension Convolution in the direction of price (CNN-GRU).

The MLP model can extract features from candlesticks without considering the temporal relationship among candles. In contrast, the other four models not only pay attention to the structure of each candlestick, but they also consider the temporal relationships. In this section, we explain the detailed architecture of each feature extractor. We will compare the performance of these DNNs as the feature extractor for the DQN later. 

Before we dive into each model's description, we need to explain different inputs to these models. Our inputs partition into two categories:
\begin{enumerate}
	\item \textbf{Vanilla}:
	
	The OHLC prices without any change. This kind of input contains only the representation of candle $c_t$ at time-step $t$.

	\item \textbf{Windowed}:
	
	A series of candlesticks with size $w$ are grouped together to form a window of candles $W = \{c_{(t - w)}, c_{(t-w+1)}, \dots, c_t\}$ at time step t.
\end{enumerate}

All models use the windowed input type, but the raw OHLC prices are only for MLP encoder and DQN without any encoder models.

\subsubsection{MLP}

The MLP model is a NN with only one hidden layer. In order to regularize the outputs of layers, we used Batch Normalization after the hidden layer. The dimensions of layers are $InputSize \times 128, BatchNormalization(128), 128 \times FeatureVectorSize$. The MLP architecture takes both types of inputs. If the raw OHLC is used, then $InputSize = 4$, and in case the input type is windowed, then the $InputSize$ would be equal to the size of the window. 

\subsubsection{GRU}
\label{sec:gru}

The Gated Recurrent Unit is a recurrent neural network exerted on extracting features from time-series data. This model's input type is the \textit{Windowed} input, which contains a sequence of candles at each time step. The role of GRU here is to extract features from each candlestick while considering the history in each window. The architecture of the GRU model is represented in \ref{fig:gru-arch}.

\subsubsection{CNN}
\label{sec:cnn}

The Convolutional Neural Networks has been widely used in image processing to extract deep features from images, and also it is applied in signal processing for analyzing signals. CNN has a kernel which can move in one, two, or higher directions and extract features from multi-dimensional data. Here, our input is the OHLC prices windowed to form time-series data. Thus, we have a 2-dimensional input, the first of which is the price(candles), and the second one is time. The input channel size is the size of each candle vector (i.e., 4 for OHLC), and the kernel size is 3 in the direction of time (i.e., w the window size). This architecture is shown in \ref{fig:cnn-arch}.

\subsubsection{CNN-GRU}
\label{sec:cnn-gru}

\begin{figure*}
	\centering
	\begin{subfigure}[b]{0.5\textwidth}
		\includegraphics[width=0.8\textwidth]{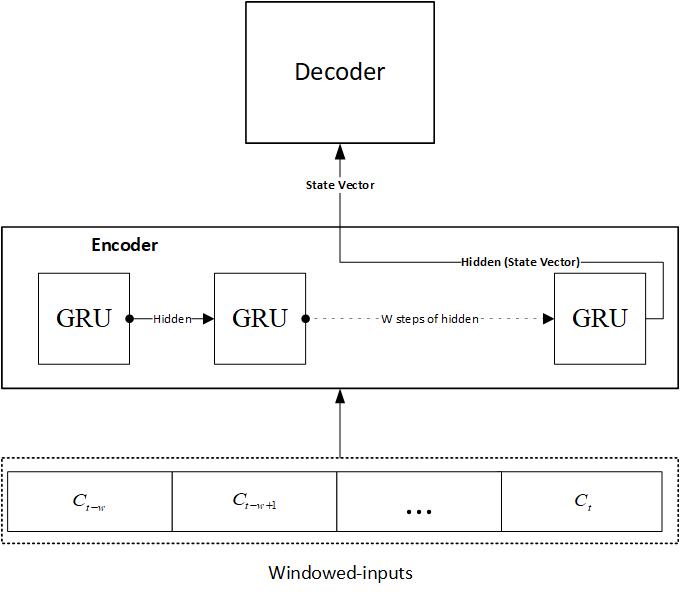}
		\caption{GRU model architecture}
		\label{fig:gru-arch}
	\end{subfigure}%
	\begin{subfigure}[b]{0.5\textwidth}
		\includegraphics[width=0.9\textwidth]{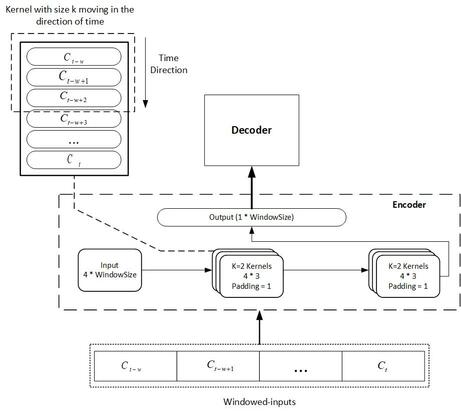}
		\caption{CNN model architecture}
		\label{fig:cnn-arch}
	\end{subfigure}
	\\
	\begin{subfigure}[b]{0.5\textwidth}
		\includegraphics[width=0.9\textwidth]{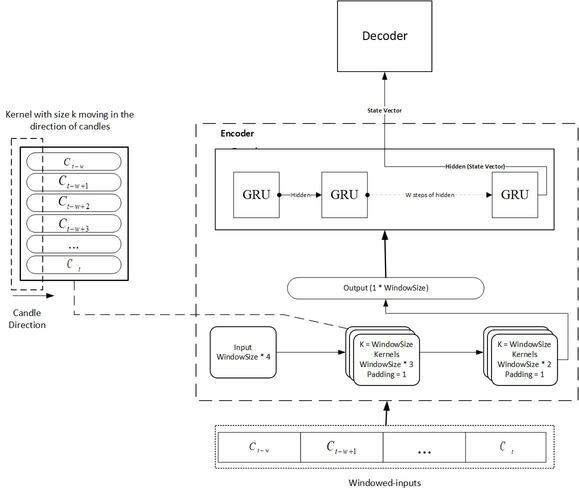}
		\caption{CNN-GRU model architecture}
		\label{fig:cnn-gru-arch}
	\end{subfigure}
	\caption{Architecture of different models proposed to use as the encoder}
\end{figure*}

The combination of CNN and GRU model is proposed here, where the CNN model's kernel moves in the direction of candles and extracts candlestick features. Then it outputs a sequence of features from the windowed input to the GRU model. The GRU model here is responsible for the input's temporal behavior, where it takes the candlesticks' features in sequence from the CNN, and extracts temporal features. The details of this architecture are represented in \ref{fig:cnn-gru-arch}.

\section{Experimental Results}
\subsection{Dataset}
All the models are tested on real-world financial data, including stocks and crypto-currencies. Data is chosen to be varied in the behavior like the bullish trend, bearish trend, and side markets. Furthermore, the data length is chosen to be 20 years with the last five years as test (trading) data, ten years with the last two years as test data, and six years (BTC/USD) with the last two years as test data. The interval of candlesticks in all data is chosen to be daily. All data used in this work are available on \textit{Yahoo Finance} and \textit{Google Finance}. The summary of the datasets is represented in \ref{tab:datasets}. 

\begin{table}[htb]
	\centering
	\caption{Data used along with train-test split dates}
	\begin{tabular}{|c||c|c|c|}
		\hline
		Data & Begin Date & Split Point & End Date \\
		\hline\hline
		GOOGL & 2010/01/01 & 2018/01/01 & 2020/08/25\\ 
		AAPL & 2010/01/01 & 2018/01/01  & 2020/08/25\\
		AAL &‌ 2010/01/01 & 2018/01/01  & 2020/08/25\\
		BTC-USD & 2014/09/17 & 2018/01/01 & 2020/08/26\\  
		KSS & 1999/01/01 & 2018/01/01 & 2020/08/24\\  
		GE & ‌2000/01/01 & 2015/01/01 & 2020/08/24\\
		HSI & ‌‌2000/01/01 & 2015/01/01 & 2020/08/24\\
		\hline    
	\end{tabular}
	\label{tab:datasets}	
\end{table}

\begin{figure*}
	\centering
	\begin{subfigure}{0.5\textwidth}
		\includegraphics[width=\linewidth, height=0.20\textheight]{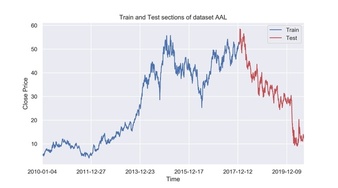}
		\caption{Price history of AAL stock used to train and test the model.}
	\end{subfigure}%
	\begin{subfigure}{0.5\textwidth}
		\includegraphics[width=\linewidth, height=0.20\textheight]{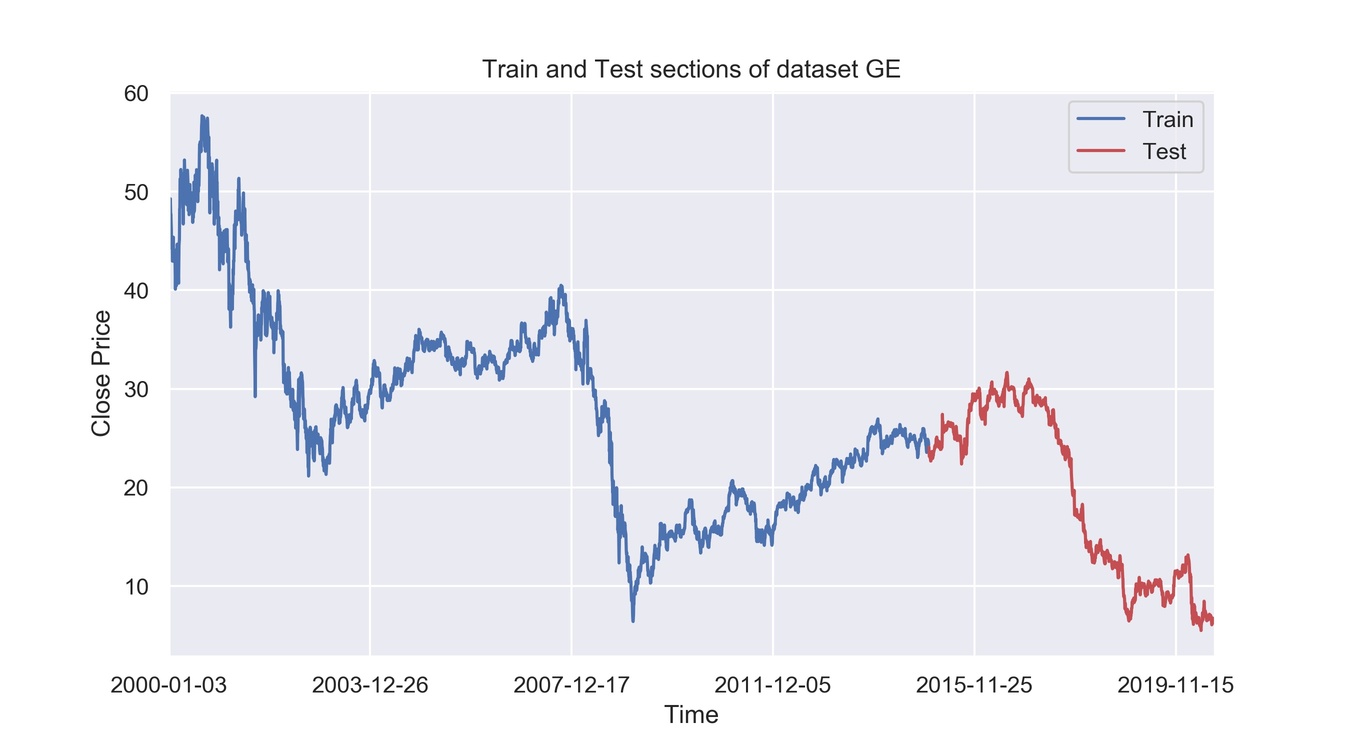}
		\caption{Price history of GE stock used to train and test the model.}
	\end{subfigure}
	\\
	\begin{subfigure}{0.5\textwidth}
		\includegraphics[width=\linewidth, height=0.20\textheight]{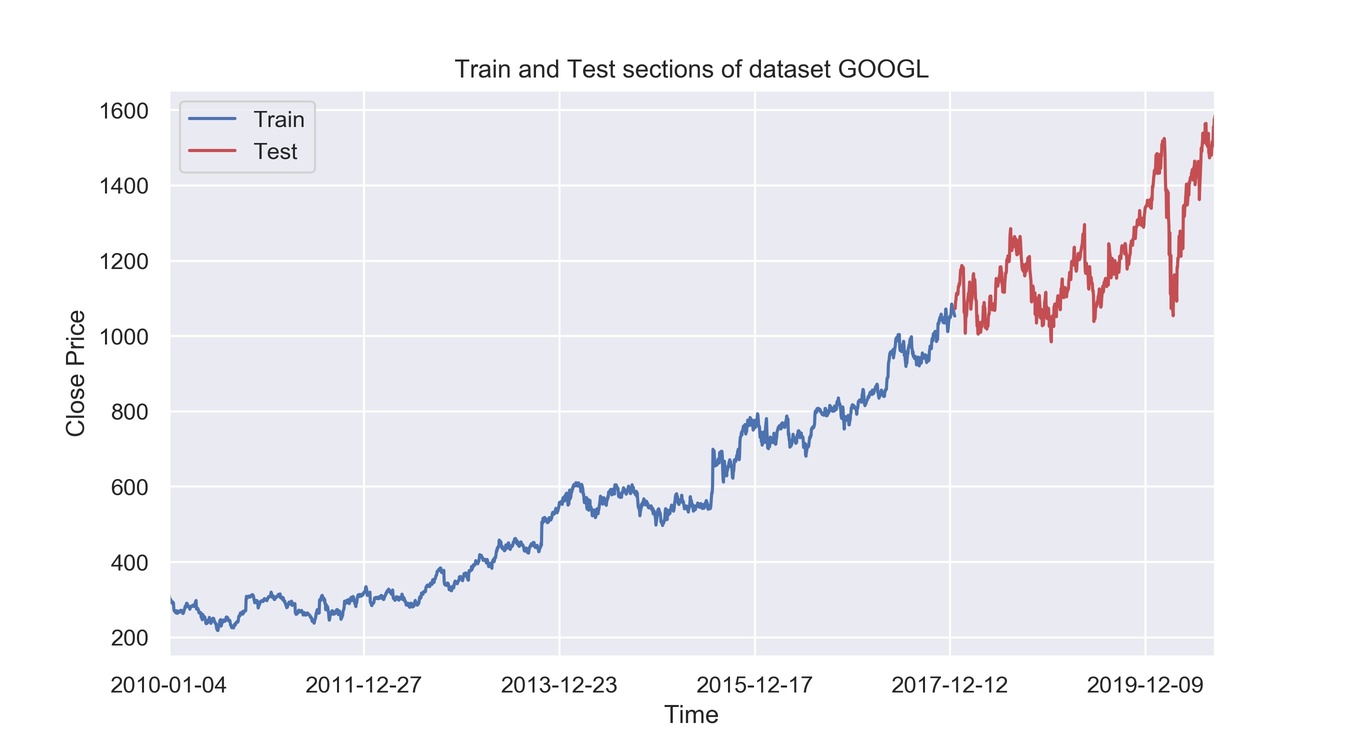}
		\caption{Price history of GOOGL stock used to train and test the model.}
	\end{subfigure}%
	\begin{subfigure}{0.5\textwidth}
		\includegraphics[width=\linewidth, height=0.20\textheight]{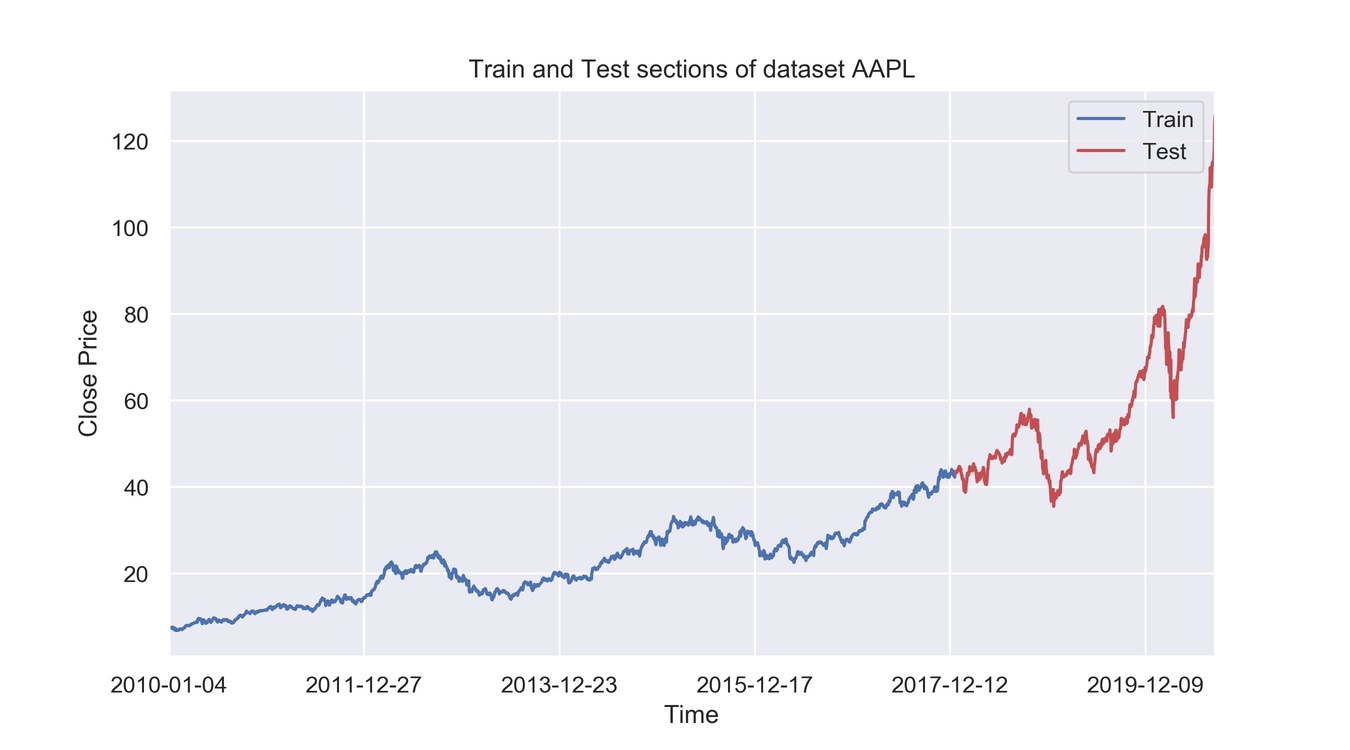}
		\caption{Price history of AAPL stock used to train and test the model.}
	\end{subfigure}
	\\
	\begin{subfigure}{0.5\textwidth}
		\includegraphics[width=\linewidth, height=0.20\textheight]{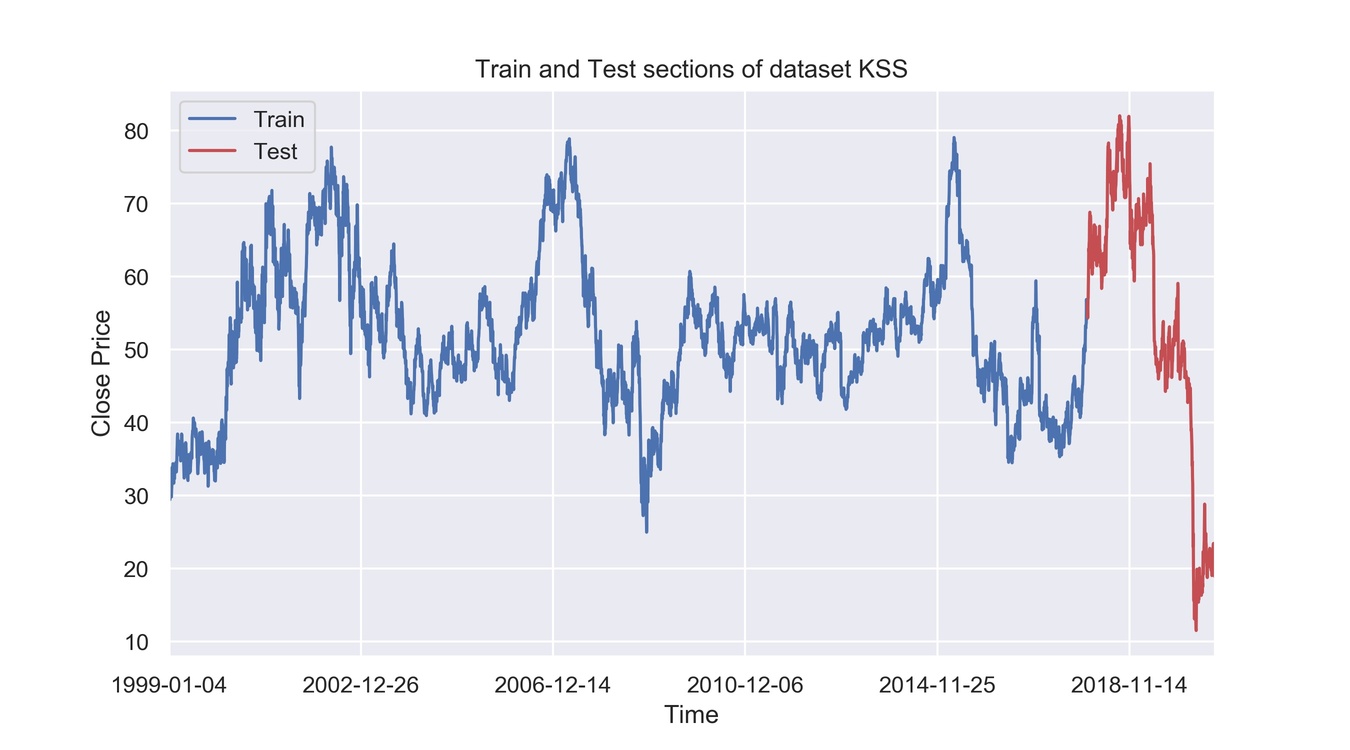}
		\caption{Price history of KSS stock used to train and test the model.}
	\end{subfigure}%
	\begin{subfigure}{0.5\textwidth}
		\includegraphics[width=\linewidth, height=0.20\textheight]{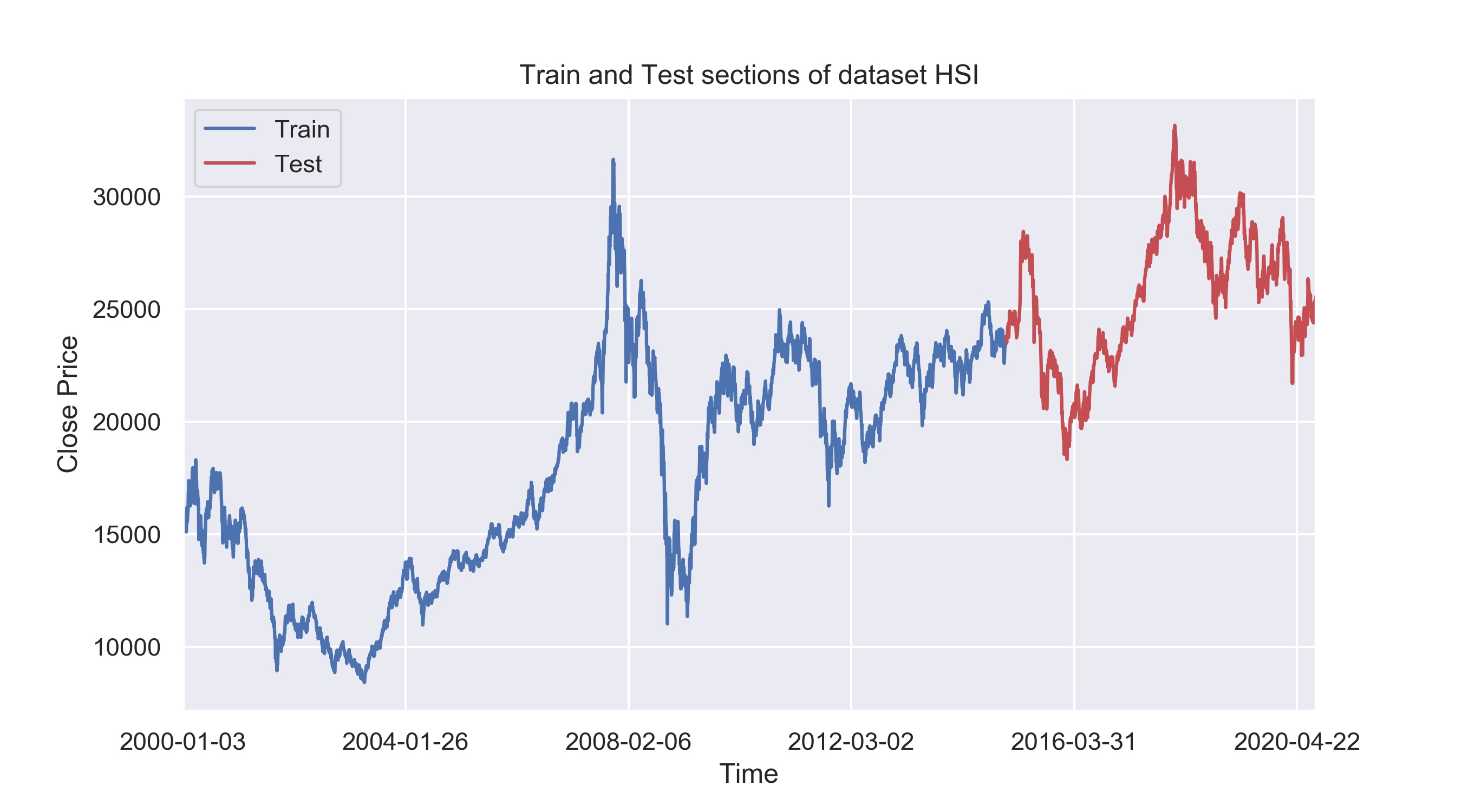}
		\caption{Price history of HSI stock used to train and test the model.}
	\end{subfigure}
	\\
	\begin{subfigure}{0.5\textwidth}
		\includegraphics[width=\linewidth, height=0.20\textheight]{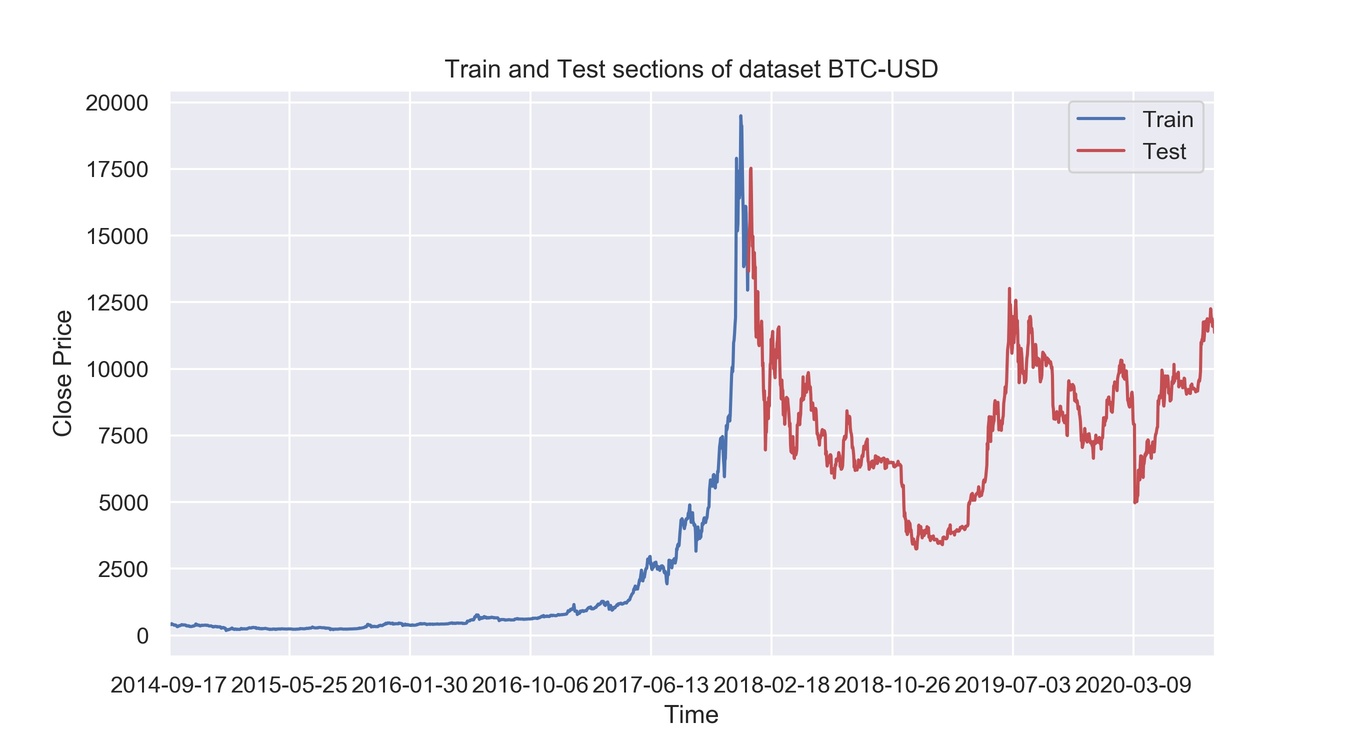}
		\caption{Price history of BTC/USD stock used to train and test the model.}
	\end{subfigure}
	\\
	\caption{Price histories used to test the models. The blue sections are used for training and the red parts are used as testing sets.}
	\label{fig:datasets}
\end{figure*}

\ref{fig:datasets} shows the condition of each dataset in different periods. The AAL data is bullish on the training-set and bearish on the test-set, market GE is both bearish on the train and test sets, AAPL and GOOGL are both bullish; KSS and HSI are examples of volatile markets, and BTC/USD is side on the test-set. These datasets are selected to measure the flexibility of different models in different market conditions. A robust model can generalize its performance to provide a proper strategy behaving profitable on the test-set. 

\subsection{Evaluation Metrics}
The trading strategy proposed by each model is evaluated from three perspectives:
\begin{enumerate}
	\item How profitable is the proposed strategy
	\item What is the risk of the proposed strategy
	\item The effect of hyper-parameters(e.g. window size) in proposing a strategy for each asset.
\end{enumerate}

The metrics are mentioned and described in detail as follows.

\subsubsection{Profit curve}
This is a qualitative metric showing the percentage of profit concerning the initial investment. At each point of time $t$, if the current wealth $W_t$ and the initial investment is $W_0$m then the percentage of the profit at each time step is calculated using \eqref{eq:profit}.
\begin{equation}
	\%Rate_t = \frac{W_t - W_0}{W_0} \times 100
	\label{eq:profit}
\end{equation}
The profit curve compares the $\%Rate$ of profit for each model at different time steps. 

\subsubsection{Arithmetic Return}
This metric is the sum of the rate of increase or decrease in the current investment due to the decisions made by the model (Buy, Sell, None). The rate of wealth change at the current time-step if the model has already invested the money (not sold before) is as in \eqref{eq:arithmetic}.

\begin{equation}
	AR_t = \frac{W_t - W_{t - 1}}{W_{t - 1}}
	\label{eq:arithmetic}
\end{equation}
Using the \eqref{eq:arithmetic}, we can calculate the arithmetic return in \eqref{eq:arithmeticreturn}.

\begin{equation}
	AR = \sum_{t = 1}^{T} AR_t
	\label{eq:arithmeticreturn}
\end{equation}

which shows the cumulative return at each time step.

\subsubsection{Time Weighted Return}
The amount of return in different periods are not independent of each other. In other words, when the amount of loss is significant at one time, then the capital would be significantly lower to invest afterward. For this purpose, we use \textit{Time Weighted Return (TWR)} which is calculated in \eqref{eq:time-weighted}.

\begin{equation}
	TWR = (\prod_{i=1}^{n}(x_i + 1))^\frac{1}{n} - 1
	\label{eq:time-weighted}
\end{equation}

To avoid negative values, we add 1 to all the return values, then we remove 1 from the result.

\subsubsection{Daily Return Variance}
This metric is the variance of daily arithmetic returns.
\begin{equation}
	RV = \frac{\sum_{t = 1}^{T}(AR_t - \overline{AR})^2}{T - 1}
\end{equation}
where $\overline{AR}$ is the average arithmetic return and $AR_t$ is the arithmetic return at time $t$.

\subsubsection{Total Return}
It is the percentage of the increase in the capital during trading time. Total Return is calculated in \eqref{eq:ttlret} where $W_0$ and $W_T$ are the initial and final wealth, respectively.

\begin{equation}
	TR = \frac{W_T - W_0}{W_0} \time 100 
	\label{eq:ttlret}
\end{equation}

\subsubsection{Value at Risk}
	The value at risk ($VaR$) is a metric to measure the quality level of financial risk within a portfolio during a specific period of time. $VaR$ typically is measured with a confidence ratio $1 - \alpha$ (e.g., with a confidence level of 95\% where $\alpha = 5$) and measures the maximum amount of loss in the worst situation with confidence $1 - \alpha$ in the corresponding time period. The higher the value of the $VaR_\alpha$ (i.e., the absolute value of $VaR_\alpha$) with a fixed value of $\alpha$, the higher the level of the portfolio's financial risk. 
	
	There exist two main approaches to compute $VaR_\alpha$: 1) using the closed-form which assumes the probability distribution of the daily returns of the portfolio follows a Normal standard distribution, 2) using the historical estimation method, which is a non-parametric method and assumes no prior knowledge about the portfolio's daily returns. In this paper, we used the closed-form method. 
	
	To calculate $VaR_{\alpha}$, 
%
	we used Monte Carlo simulation by developing a model for future stock price returns and running multiple hypothetical trials through the model. The mean $\mu$ and standard deviation $\sigma$ of the returns are calculated, then 1000 simulations run to generate random outputs with a normal distribution $N(\mu, \sigma)$. Then the $\alpha$ percent lowest value of the outputs is selected and reported as $VaR_\alpha$.

\subsubsection{Daily Return Volatility}
The volatility of the daily returns evaluates the risk level of trading rules by calculating daily returns' standard deviation. This metric is calculated for each strategy using \eqref{eq:volatility}, where $\overline{AR}$ is the average daily arithmetic return, and AR is the daily arithmetic return.

\begin{equation}
	\sigma_p = \sqrt{\frac{\Sigma_{i=1}^T (AR_i - \overline{AR})^2}{T-1}} 
	\label{eq:volatility}
\end{equation}

\subsubsection{Sharpe Ratio}
The Sharpe ratio (SR) was proposed first by Sharpe et al. \cite{sharpe1994sharpe} to measure the reward-to-variability ratio of the mutual funds. This metric displays the average return earned in excess of the risk-free rate per unit total risk and is computed here by \eqref{eq:SR} in which $R_f$ is the return of the \textit{risk-free} asset, and $E\{R_p\}$ is the expected value of the portfolio value. Here we assumed that $R_f=0$.

\begin{equation}
	SR = \frac{E\{R_p\} - R_f}{\sigma_p}
	\label{eq:SR}
\end{equation}

\subsubsection{Window Size heat-map}
This diagram illustrates the impact of window-size in extracting appropriate patterns from the input candlesticks for each asset, which is reflected as the total profit earned by the agent corresponding to each window size. 

\subsubsection{Decision curve} 
In this curve, the trading signals to trade each asset are demonstrated over that asset's raw price curve. This chart gives insight into the quality of decision making power of each model on each financial asset.

\subsection{Experimental setup}
All the models are implemented using \textit{Pytorch} library in Python. In order to optimize the models, we used Adam optimizer. The mini-batch training is also conducted using a batch size of 10, and the replay memory size is set to 20. The only regularization used in the experiments in the policy and target networks is the \textit{Batch Normalization}. The transaction cost is set to zero during the training process; however, it may be non-zero during the evaluation.

\subsection{Performance evaluation of the models}
In this section, the overall performance of the different models, along with different input types for the MLP and DQN models (i.e., raw OHLC and Windowed candles) are compared using profit curves and other risk and profit evaluation metrics.

\begin{figure*}
	\centering
	\begin{subfigure}{0.5\textwidth}
		\includegraphics[width=\linewidth, height=0.20\textheight]{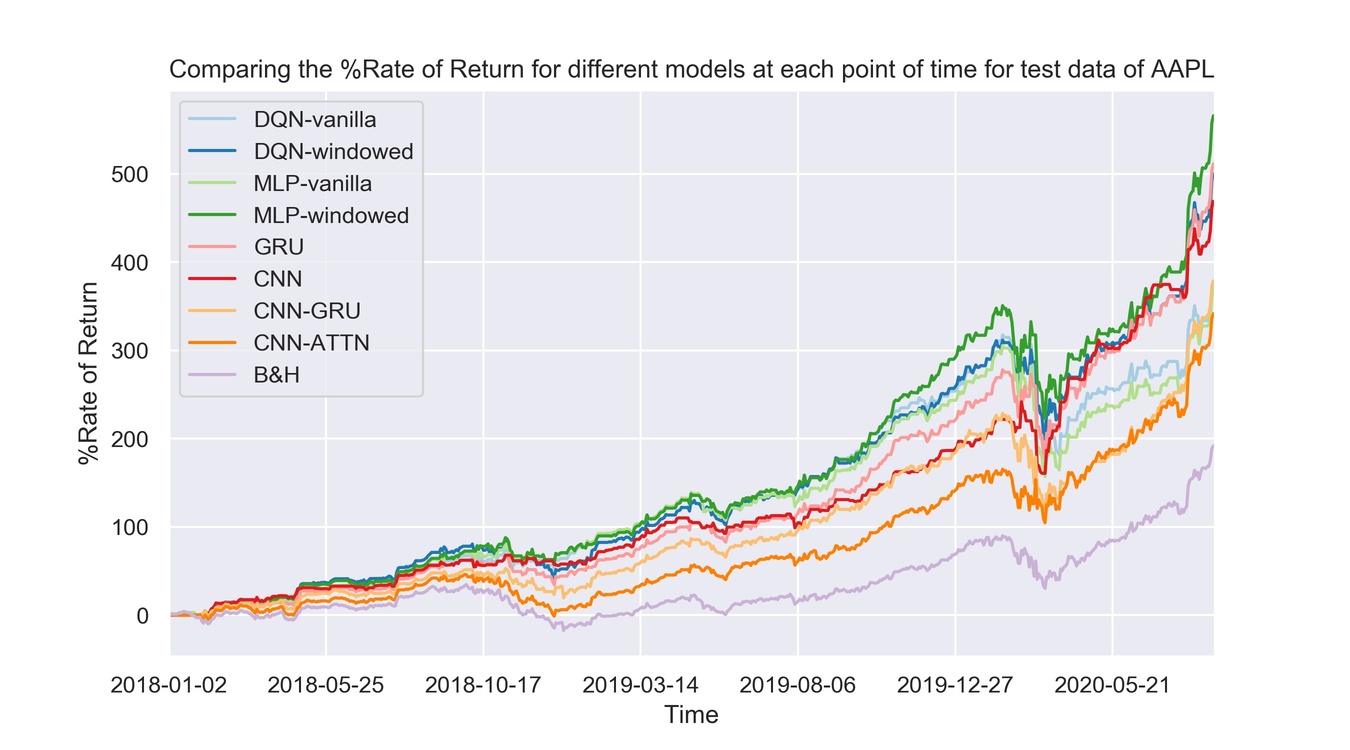}
		\caption{Performance of different models on AAPL}
	\end{subfigure}%
	\begin{subfigure}{0.5\textwidth}
		\includegraphics[width=\linewidth, height=0.20\textheight]{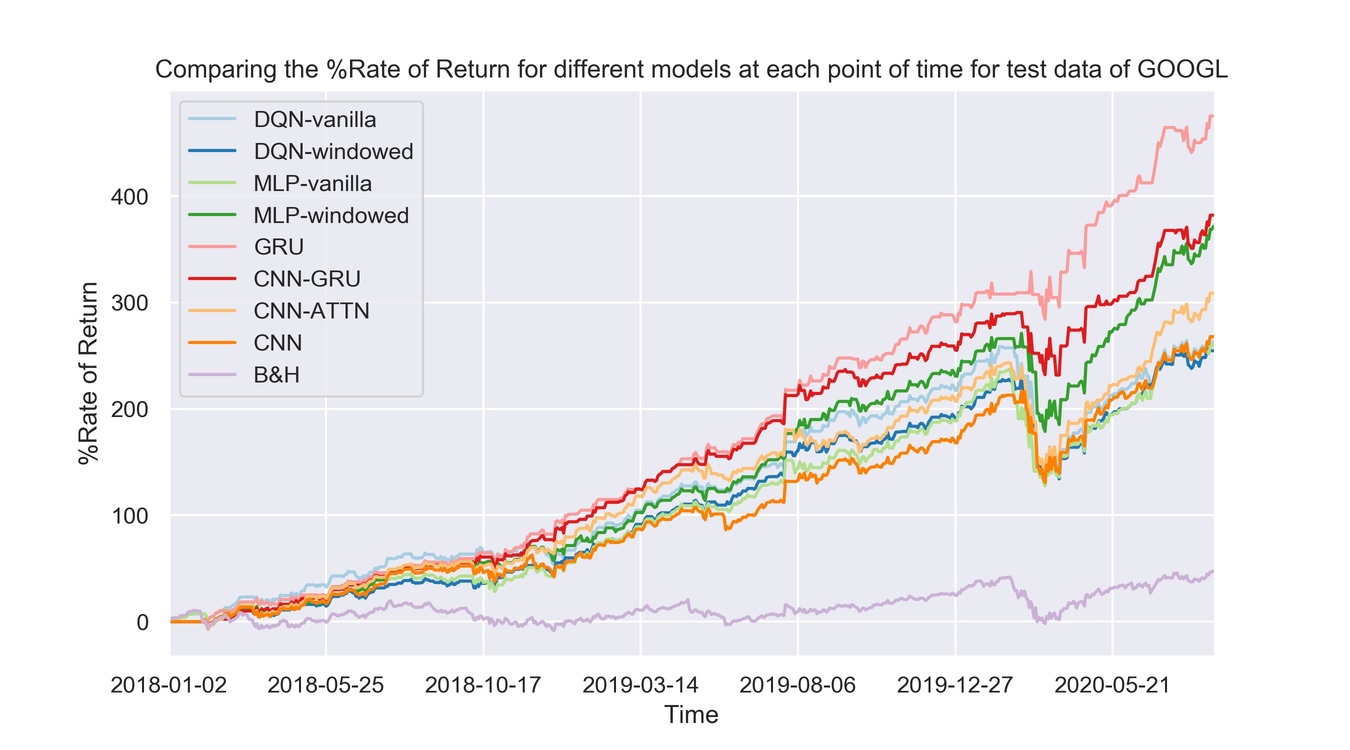}
		\caption{Performance of different models on GOOGL}
	\end{subfigure}
	\\
	\begin{subfigure}{0.5\textwidth}
		\includegraphics[width=\linewidth, height=0.20\textheight]{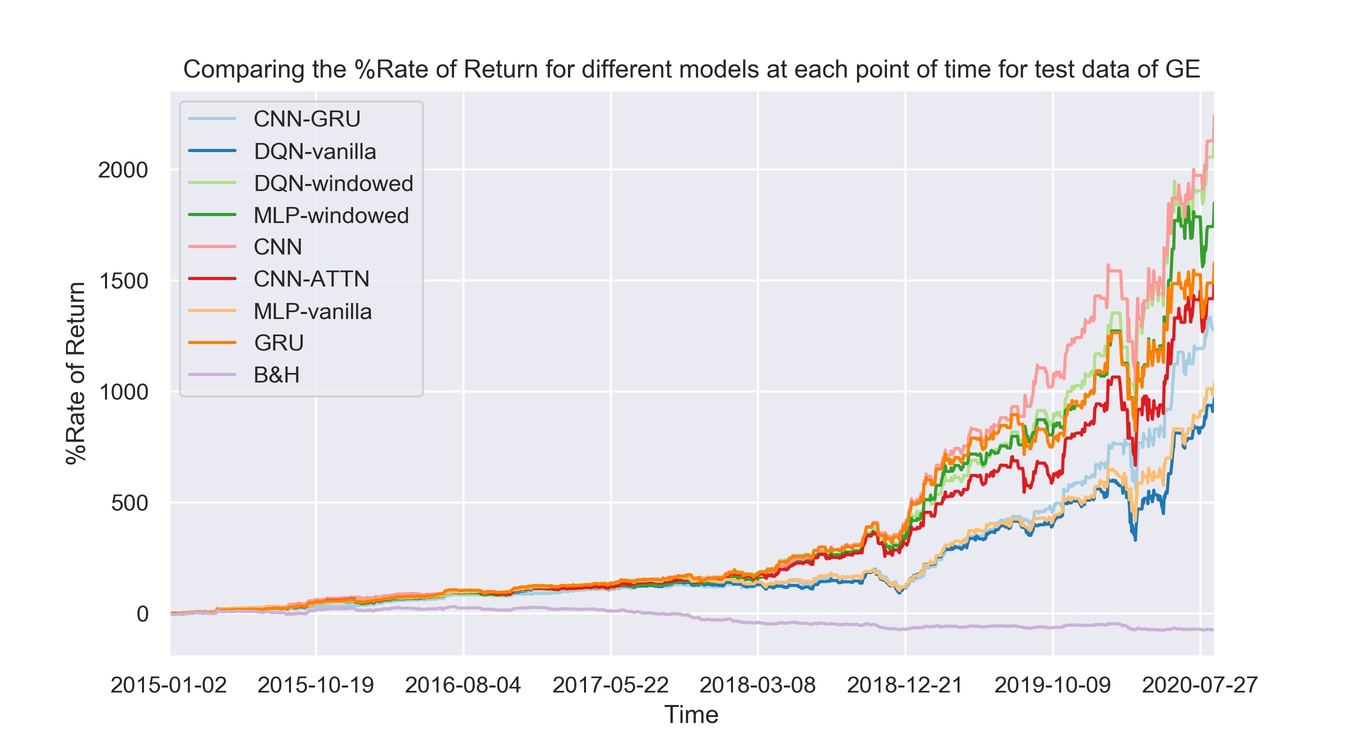}
		\caption{Performance of different models on GE}
	\end{subfigure}%
	\begin{subfigure}{0.5\textwidth}
		\includegraphics[width=\linewidth, height=0.20\textheight]{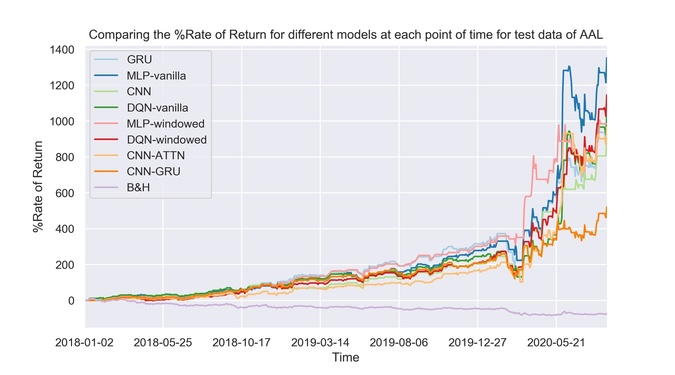}
		\caption{Performance of different models on AAL}
	\end{subfigure}
		\\
	\begin{subfigure}{0.5\textwidth}
		\includegraphics[width=\linewidth, height=0.20\textheight]{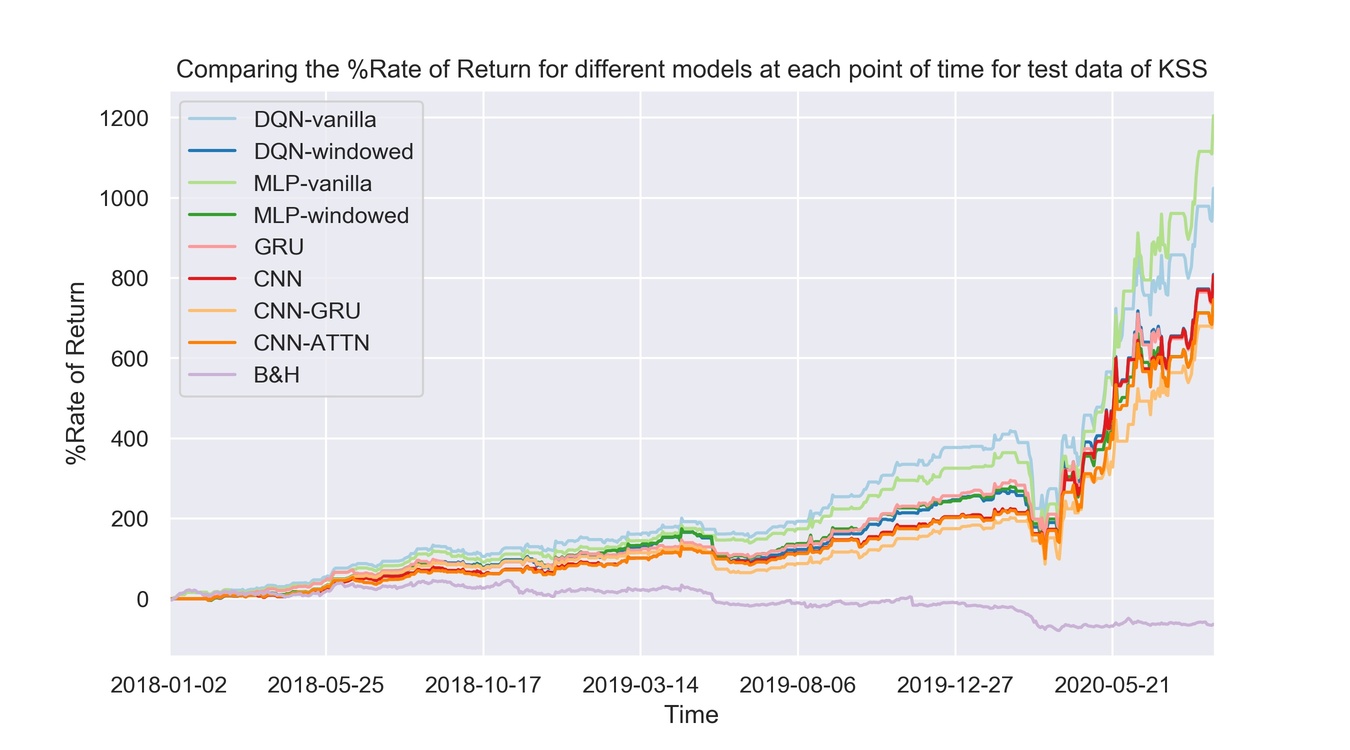}
		\caption{Performance of different models on KSS}
	\end{subfigure}%
	\begin{subfigure}{0.5\textwidth}
		\includegraphics[width=\linewidth, height=0.20\textheight]{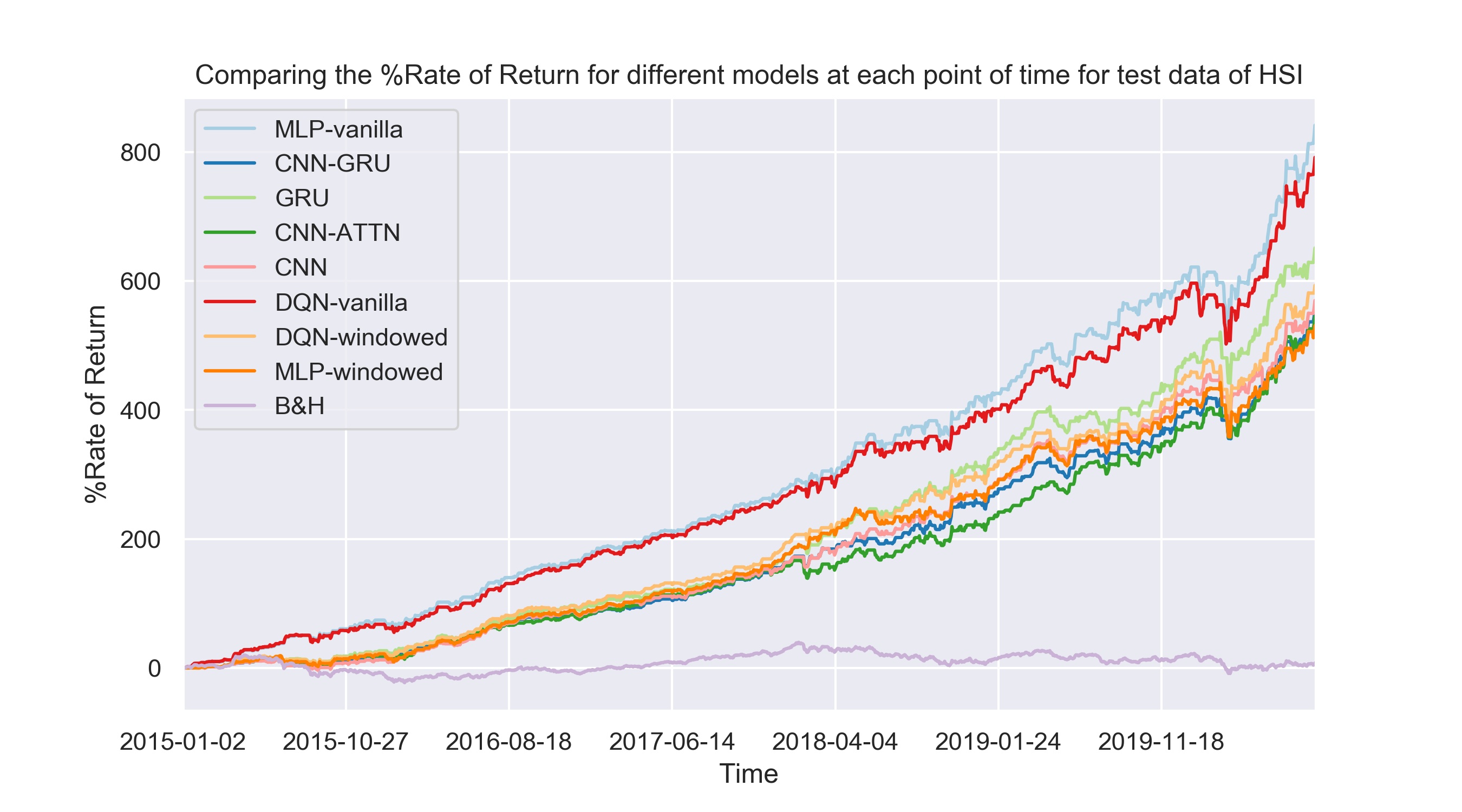}
		\caption{Performance of different models on HSI}
	\end{subfigure}
	\\
	\begin{subfigure}{0.5\textwidth}
		\includegraphics[width=\linewidth, height=0.20\textheight]{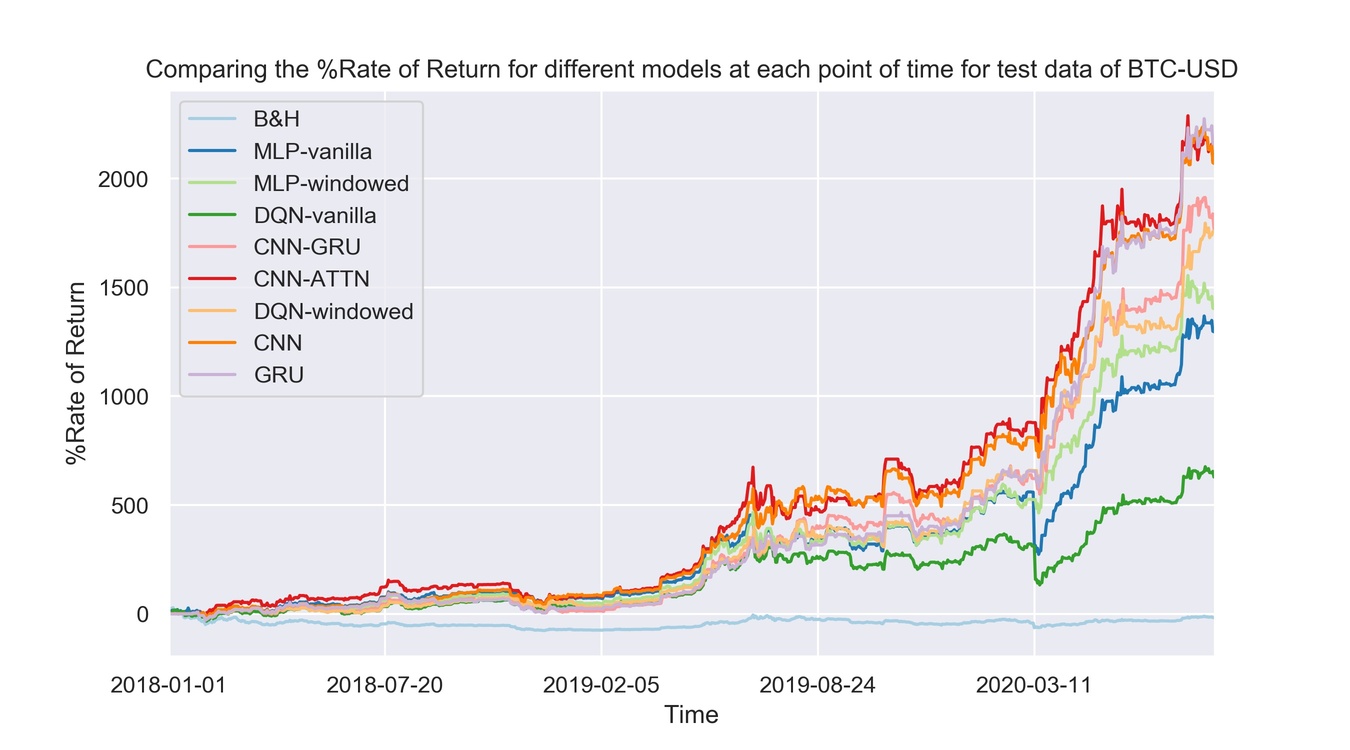}
		\caption{Performance of different models on BTC/USD}
	\end{subfigure}

	\caption{The profit curve of the models different from the viewpoint of encoder part and input type.}
	\label{fig:testset-performance}
\end{figure*}

\ref{fig:testset-performance} illustrates the profit curves of models on the test set for different datasets. DQN-vanilla is the DQN model without any encoder and with the input of raw OHLC. DQN-windowed is the DQN model with a window of candles as input. MLP-vanilla and MLP-windowed are the same as DQN-vanilla and DQN-windowed except that MLP contains an encoder part, which is an MLP model. CNN, GRU, and CNN-GRUare models with the encoder part as described in sections \ref{sec:cnn}, \ref{sec:gru}, and \ref{sec:cnn-gru}, respectively with input type as a window of candles(time series).

The general conclusions we reached from the experiments are reported in \ref{fig:testset-performance}:
\begin{itemize}
	\item 
	Stocks can be categorized into two kinds: the one in which the sequence of candlesticks have effective temporal relationships and those with few meaningful time dependencies. 
	
	\item 
	The most profitable trading strategies for data with a high level of time-dependency in their price history can be generated using windowed-input models. The BTC/USD, GOOGL, AAPL, and GE, are of this kind. The GRU, and CNN have the best performance on the BTC/USD model; The CNN, DQN-windowed, and MLP-windowed have the best performance on GE; The GRU, CNN-GRU, and MLP-windowed have the best performance on GOOGL; MLP-windowed, GRU, DQN-windowed, and CNN provided the most profitable strategies for AAPL.
	
	\item On the other hand, we have data with a low level of dependency in time among candlesticks, which leads to the models with raw OHLC input having a better performance. AAL, HSI, and KSS are among this type of data. By low level of dependency in time, we do not mean that models with time-series inputs have poor performance. Their performance is very good, but they behave a little poorly compared to models with raw OHLC input.

\end{itemize}

\ref{tab:results1} represents the details of experiments with regards to both profit and risk. One crucial point that can be inferred from the results is that as the models' total return increases, the Sharpe ratio increases correspondingly. It means that the models devise strategies in a risk-adjusted way. However, if we want to examine the results in specific, on some data, the models with the highest profitability acted riskily. For example, in AAL, the best model in total return is MLP-vanilla, but the Sharpe ratio and VaR(here we consider the absolute value of VaR) of its strategy are, respectively, lower and higher than those strategies proposed by windowed-input models. The same is true about GRU in BTC/USD, where GRU has the highest total return, but CNN provided more risk-adjusted strategies with respect to VaR and Sharpe ratio.

Another important conclusion deducted from comparing data diagrams and the models' performance is that stocks with highly volatile prices such as KSS, AAL, and HSI can best be processed by models with raw OHLC inputs, whereas other data with more stable prices are best analyzed with windowed-input models. Therefore, in order to select among feature extractors, we should pay careful attention to the type of input data. The best feature extractor for stable stocks would be windowed-input models, whereas, for highly volatile stocks, models with raw OHLC can decide and change their behavior more quickly since they only pay attention to the current candlestick, not the history of candles.

\begin{table*}[htb]
	\centering
	\caption{Performance of different models on BTC/USD, GOOGL, AAPL, KSS, and GE}
	\label{tab:results1}
	\begin{adjustbox}{width=0.8\textwidth}
		\begin{tabular}{@{} |l*{10}{|l}| @{}}
			\hline
			Agent & \rotatebox[origin=c]{90}{Arithmetic Return} & \rotatebox[origin=c]{90}{Average Daily Return} & \rotatebox[origin=c]{90}{Daily Return Variance} & \rotatebox[origin=c]{90}{Time Weighted Return} & \rotatebox[origin=c]{90}{Total Return}  & \rotatebox[origin=c]{90}{Sharpe Ratio} & \rotatebox[origin=c]{90}{Value At Risk} & \rotatebox[origin=c]{90}{Volatility} & \rotatebox[origin=c]{90}{Initial Investment} & \rotatebox[origin=c]{90}{Final Portfolio Value}\\
			\hline
			\hline
			\multicolumn{11}{c}{BTC/USD} \\
			\hline
			DQN-vanilla & 262 & 0.27 & 12.64 & 0.002 & 629 \% & 0.076 & -5.58 & 110.6 & 1000 & 7287.2 \\
			\hline
			DQN-windowed & 334 & 0.34 & 8.69 & 0.003 & 1757 \% & 0.117 & -4.51 & 91.7 & 1000 & 18567.1 \\
			\hline
			MLP-vanilla & 324 & 0.33 & 12.01 & 0.003 & 1296 \% & 0.097 & -5.37 & 107.8 & 1000 & 13959.8 \\
			\hline
			MLP-windowed & 320 & 0.33 & 10.19 & 0.003 & 1402 \% & 0.104 & -4.93 & 99.3 & 1000 & 15021.8 \\
			\hline
			\hline
			GRU & 359 & 0.37 & 9.89 & 0.003 & 2158 \% & 0.118 & -4.81 & 97.8 & 1000 & 22577.1 \\
			\hline
			CNN & 353 & 0.36 & 9.42 & 0.003 & 2069 \% & 0.119 & -4.69 & 95.5 & 1000 & 21693.6 \\	
			\hline
			CNN-GRU & 338 & 0.35 & 9.47 & 0.003 & 1770 \% & 0.114 & -4.72 & 95.7 & 1000 & 18701.5 \\
			\hline
			\hline
			\hline
			\multicolumn{11}{c}{GOOGL} \\
			\hline
			DQN-vanilla & 138 & 0.21 & 2.59 & 0.002 & 263 \% & 0.128 & -2.45 & 41.6 & 1000 & 3631.0 \\
			\hline
			DQN-windowed & 134 & 0.20 & 2.25 & 0.002 & 255 \% & 0.134 & -2.27 & 38.7 & 1000 & 3546.3 \\
			\hline
			MLP-vanilla & 135 & 0.20 & 2.60 & 0.002 & 252 \% & 0.125 & -2.45 & 41.6 & 1000 & 3520.7 \\
			\hline
			MLP-windowed & 163 & 0.24 & 2.29 & 0.002 & 371 \% & 0.162 & -2.25 & 39.0 & 1000 & 4714.4 \\
			\hline
			\hline
			GRU & 180 & 0.27 & 1.56 & 0.003 & 475 \% & 0.217 & -1.79 & 32.2 & 1000 & 5752.0 \\
			\hline
			CNN & 139 & 0.21 & 2.73 & 0.002 & 268 \% & 0.127 & -2.51 & 42.6 & 1000 & 3678.4 \\	
			\hline
			CNN-GRU & ‌163 & 0.25 & 1.75 & 0.002 & 382 \% & 0.185 & -1.93 & 34.1 & 1000 & 4819.9 \\
			\hline
			\hline
			\hline
			\multicolumn{11}{c}{AAPL} \\
			\hline
			DQN-vanilla & 166 & 0.25 & 3.07 & 0.002 & 372 \% & 0.142 & -2.63 & 45.2 & 1000 & 4722.5 \\
			\hline
			DQN-windowed & 190 & 0.29 & 3.22 & 0.003 & 500 \% & 0.159 & -2.67 & 46.3 & 1000 & 5997.1 \\
			\hline
			MLP-vanilla & 165 & 0.25 & 3.16 & 0.002 & 366 \% & 0.139 & -2.68 & 45.9 & 1000 & 4657.1 \\
			\hline
			MLP-windowed & 200 & 0.30 & 3.03 & 0.003 & 566 \% & 0.172 & -2.57 & 44.9 & 1000 & 6658.3 \\
			\hline
			\hline
			GRU & 191 & 0.29 & 2.99 & 0.003 & 511 \% & 0.166 & -2.56 & 44.6 & 1000 & 6112.8 \\
			\hline
			CNN & 181 & 0.27 & 2.07 & 0.003 & 469 \% & 0.189 & -2.10 & 37.1 & 1000 & 5688.3 \\	
			\hline
			CNN-GRU & 170 & 0.26 & 4.03 & 0.002 & 379 \% & 0.127 & -3.05 & 51.8 & 1000 & 4786.2 \\
			\hline
			\hline
			\hline
			\multicolumn{11}{c}{KSS} \\
			\hline	
			DQN-vanilla & 272 & 0.41 & 9.25 & 0.004 & 1024 \% & 0.134 & -4.60 & 78.5 & 1000 & 11236.7 \\
			\hline
			DQN-windowed & 251 & 0.38 & 9.38 & 0.003 & 809 \% & 0.123 & -4.67 & 79.0 & 1000 & 9088.3 \\
			\hline
			MLP-vanilla & 287 & 0.43 & 9.14 & 0.004 & 1205 \% & 0.143 & -4.55 & 78.0 & 1000 & 13048.2 \\
			\hline
			MLP-windowed & 242 & 0.36 & 8.93 & 0.003 & 747 \% & 0.122 & -4.56 & 77.1 & 1000 & 8467.0 \\
			\hline
			\hline
			GRU & 248 & 0.37 & 8.84 & 0.003 & 801 \% & 0.125 & -4.52 & 76.7 & 1000 & 9005.2 \\
			\hline
			CNN & 250 & 0.37 & 9.10 & 0.003 & 806 \% & 0.124 & -4.59 & 77.8 & 1000 & 9055.3 \\	
			\hline
			CNN-GRU & 242 & 0.36 & 9.19 & 0.003 & 737 \% & 0.120 & -4.63 & 78.3 & 1000 & 8369.1 \\
			\hline
			\hline
			\hline
			\multicolumn{11}{c}{GE} \\
			\hline
			DQN-vanilla & 260 & 0.18 & 3.25 & 0.002 & 967 \% & 0.101 & -2.79 & 68.0 & 1000 & 10673.4 \\
			\hline
			DQN-windowed & 333 & 0.23 & 2.87 & 0.002 & 2179 \% & 0.138 & -2.56 & 63.9 & 1000 & 22788.3 \\
			\hline
			MLP-vanilla & 264 & 0.19 & 2.87 & 0.002 & 1044 \% & 0.110 & -2.61 & 63.9 & 1000 & 11442.2 \\
			\hline
			MLP-windowed & 317 & 0.22 & 2.89 & 0.002 & 1848 \% & 0.131 & -2.58 & 64.1 & 1000 & 19482.2 \\
			\hline
			\hline
			GRU & 304 & 0.21 & 3.12 & 0.002 & 1580 \% & 0.121 & -2.69 & 66.5 & 1000 & 16795.4 \\
			\hline
			CNN & 335 & 0.24 & 2.74 & 0.002 & 2242 \% & 0.142 & -2.49 & 62.3 & 1000 & 23416.3 \\	
			\hline
			CNN-GRU & 283 & 0.20 & 2.91 & 0.002 & 1278 \% & 0.117 & -2.61 & 64.3 & 1000 & 13779.3 \\
			\hline
			\hline
		\end{tabular}
	\end{adjustbox} 
\end{table*}

\begin{table*}[t]
	\centering
	\begin{adjustbox}{width=0.8\textwidth}
		\begin{tabular}{@{} |l*{10}{|l}| @{}}
			\hline
			Agent & \rotatebox[origin=c]{90}{Arithmetic Return} & \rotatebox[origin=c]{90}{Average Daily Return} & \rotatebox[origin=c]{90}{Daily Return Variance} & \rotatebox[origin=c]{90}{Time Weighted Return} & \rotatebox[origin=c]{90}{Total Return}  & \rotatebox[origin=c]{90}{Sharpe Ratio} & \rotatebox[origin=c]{90}{Value At Risk} & \rotatebox[origin=c]{90}{Volatility} & \rotatebox[origin=c]{90}{Initial Investment} & \rotatebox[origin=c]{90}{Final Portfolio Value}\\
			\hline
			\hline
			\multicolumn{11}{c}{HSI} \\
			\hline
			DQN-vanilla & 224 & 0.16 & 0.78 & 0.002 & 792 \% & 0.182 & -1.30 & 33.0 & 1000 & 8915.1 \\
			\hline
			DQN-windowed & 199 & 0.14 & 0.72 & 0.001 & 593 \% & 0.169 & -1.25 & 31.6 & 1000 & 6930.6 \\
			\hline
			MLP-vanilla & 230 & 0.17 & 0.76 & 0.002 & 841 \% & 0.189 & -1.27 & 32.6 & 1000 & 9412.9 \\
			\hline
			MLP-windowed & 190 & 0.14 & 0.82 & 0.001 & 531 \% & 0.151 & -1.35 & 33.7 & 1000 & 6306.6 \\
			\hline
			\hline
			GRU & 207 & 0.15 & 0.81 & 0.001 & 651 \% & 0.166 & -1.33 & 33.6 & 1000 & 7510.0 \\
			\hline
			CNN & 196 & 0.14 & 0.80 & 0.001 & 570 \% & 0.157 & -1.33 & 33.4 & 1000 & 6696.0 \\	
			\hline
			CNN-GRU & 193 & 0.14 & 0.72 & 0.001 & 556 \% & 0.164 & -1.26 & 31.6 & 1000 & 6560.5 \\
			\hline
			\hline
			\hline
			\multicolumn{11}{c}{AAL} \\
			\hline
			DQN-vanilla & 280 & 0.42 & 12.07 & 0.004 & 1030 \% & 0.121 & -5.30 & 89.7 & 1000 & 11299.0 \\
			\hline
			DQN-windowed & 281 & 0.42 & 9.14 & 0.004 & 1145 \% & 0.140 & -4.56 & 78.0 & 1000 & 12447.2 \\
			\hline
			MLP-vanilla & 303 & 0.45 & 11.45 & 0.004 & 1352 \% & 0.134 & -5.12 & 87.3 & 1000 & 14521.4 \\
			\hline
			MLP-windowed & 261 & 0.39 & 7.03 & 0.004 & 985 \% & 0.148 & -3.98 & 68.4 & 1000 & 10845.0 \\
			\hline
			\hline
			GRU & 268 & 0.40 & 8.77 & 0.004 & 999 \% & 0.136 & -4.48 & 76.4 & 1000 & 10989.6 \\
			\hline
			CNN & 253 & 0.38 & 7.19 & 0.003 & 900 \% & 0.142 & -4.04 & 69.2 & 1000 & 9999.2 \\	
			\hline
			CNN-GRU & 212 & 0.32 & 9.36 & 0.003 & 520 \% & 0.104 & -4.72 & 79.0 & 1000 & 6198.9 \\
			\hline
			\hline
		\end{tabular}
	\end{adjustbox} 
\end{table*}

\subsection{impact of window size}
\begin{figure}[htb]
	\centering
	\includegraphics[width=0.55\textwidth]{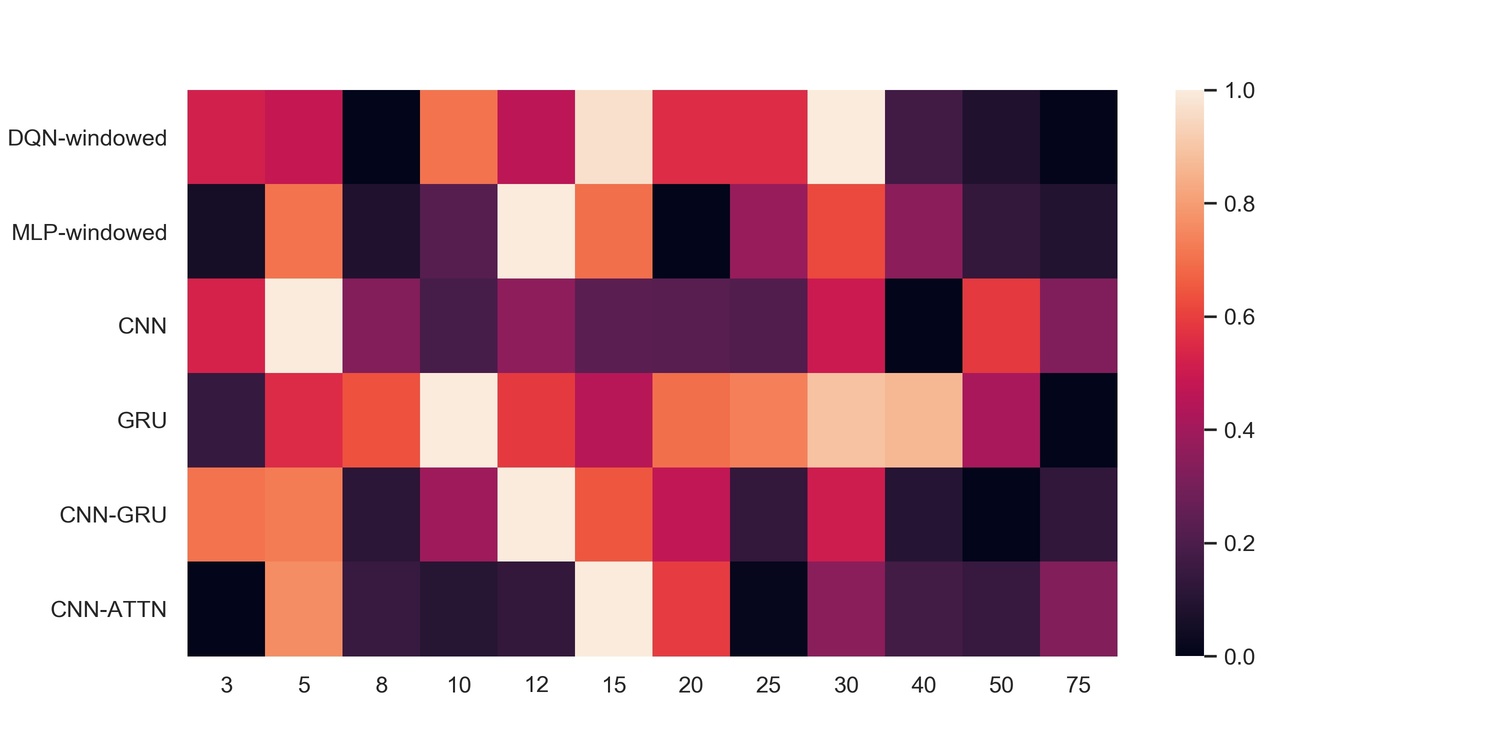}
	\caption{The heat-map generated to show the impact of different window sizes on the feature extraction by calculating total return for each window size.}
	\label{fig:window}
\end{figure}

Now that we have examined the performance of different feature extractor models, it is time to dive deeper into the temporal feature extractor concept. Feature extractors with windowed inputs can perform better on data with more stable price movements (rather than highly volatile data). Moreover, considering \ref{tab:results1} and \ref{fig:testset-performance}, each windowed-input model has its best performance varying from data to data. We want to inspect the impact of window size for each model differently using the data in which the model has its best performance. We test the performance of GRU and CNN-GRU using GOOGL; CNN, MLP-windowed, and DQN-windowed using GE. \ref{fig:window} demonstrate a heat-map showing the relationship between the window size and the normalized total profit earned by the models with windowed input. Window sizes vary from 3 to 75, and the total profit is normalized to bring between 0 and 1. Blocks having lighter colors earned higher profit than those with darker colors. As obvious from the heat-map, the number of lighter colors in the interval of 10 to 20 is more than other window sizes. Therefore, the best feature extraction using a sequence of candlesticks can be done with window sizes between 10 and 20.

\subsection{Sample Signaling}
\begin{figure*}
	\centering
	\begin{subfigure}{0.5\textwidth}
		\includegraphics[width=\linewidth, height=0.20\textheight]{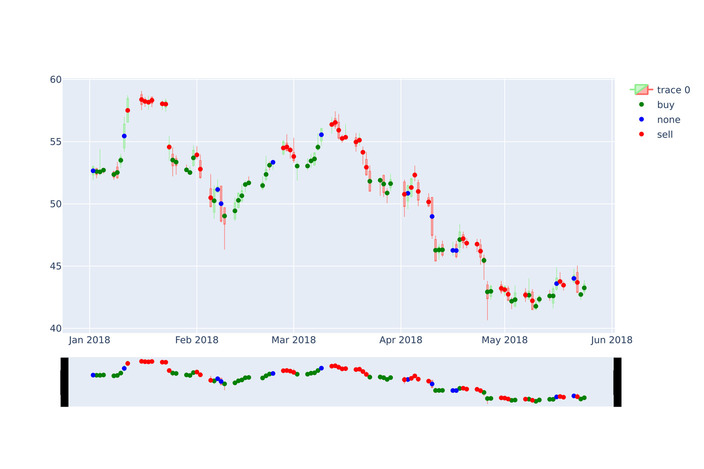}
		\caption{Trading strategy generated for AAL}
	\end{subfigure}%
	\begin{subfigure}{0.5\textwidth}
		\includegraphics[width=\linewidth, height=0.20\textheight]{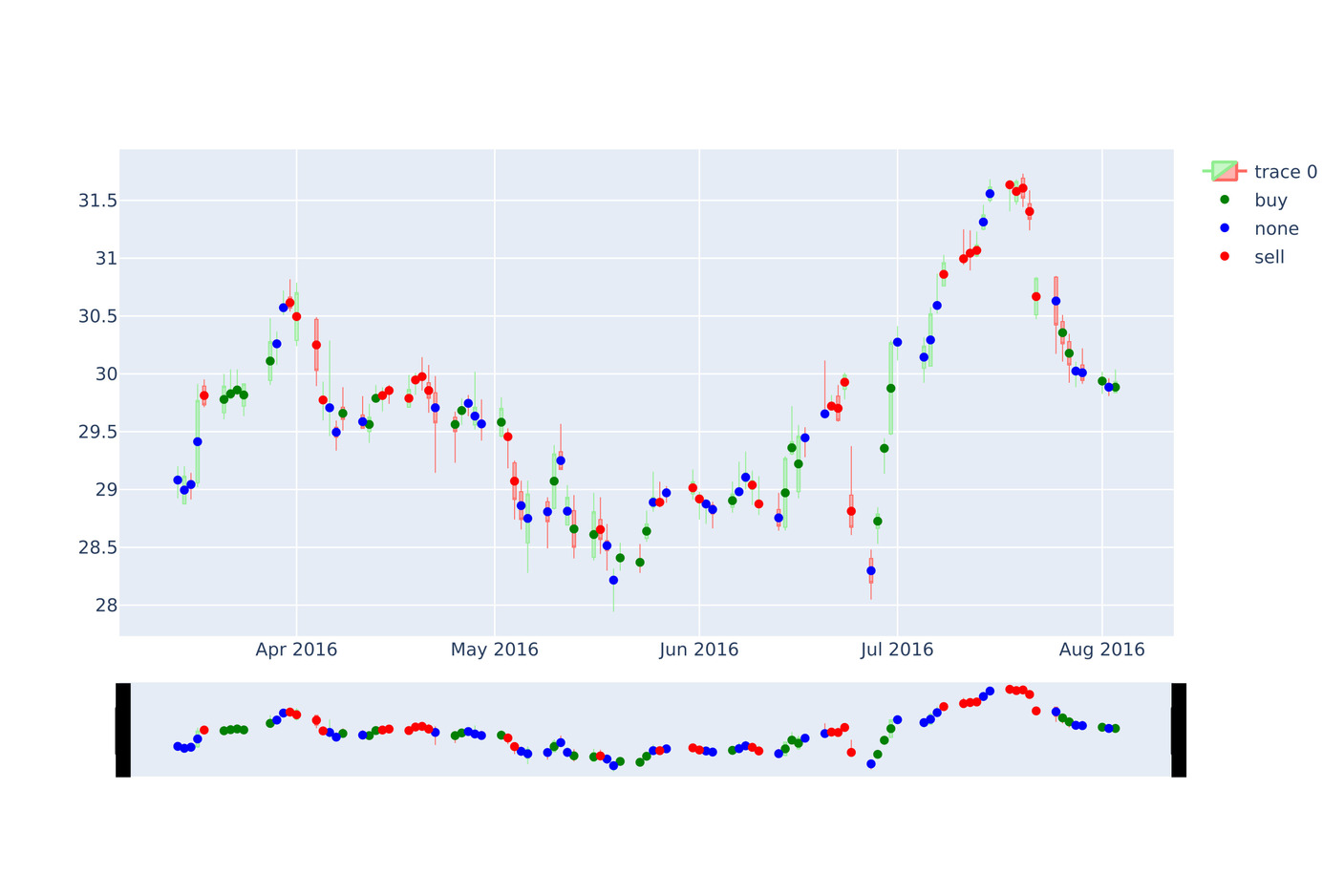}
		\caption{Trading strategy generated for GE}
	\end{subfigure}
	\\
	\begin{subfigure}{0.5\textwidth}
		\includegraphics[width=\linewidth, height=0.20\textheight]{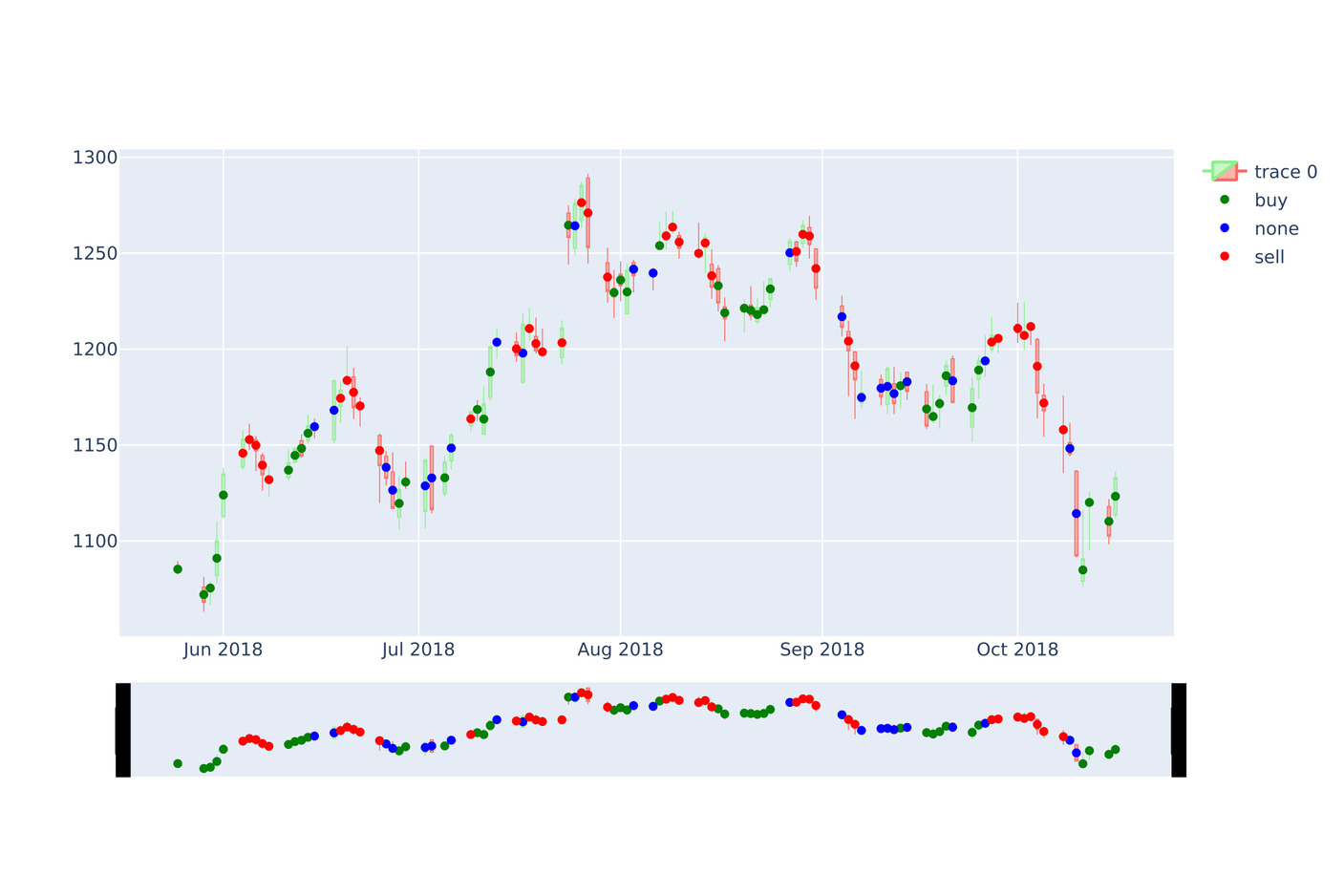}
		\caption{Trading strategy generated for GOOGL}
	\end{subfigure}%
	\begin{subfigure}{0.5\textwidth}
		\includegraphics[width=\linewidth, height=0.20\textheight]{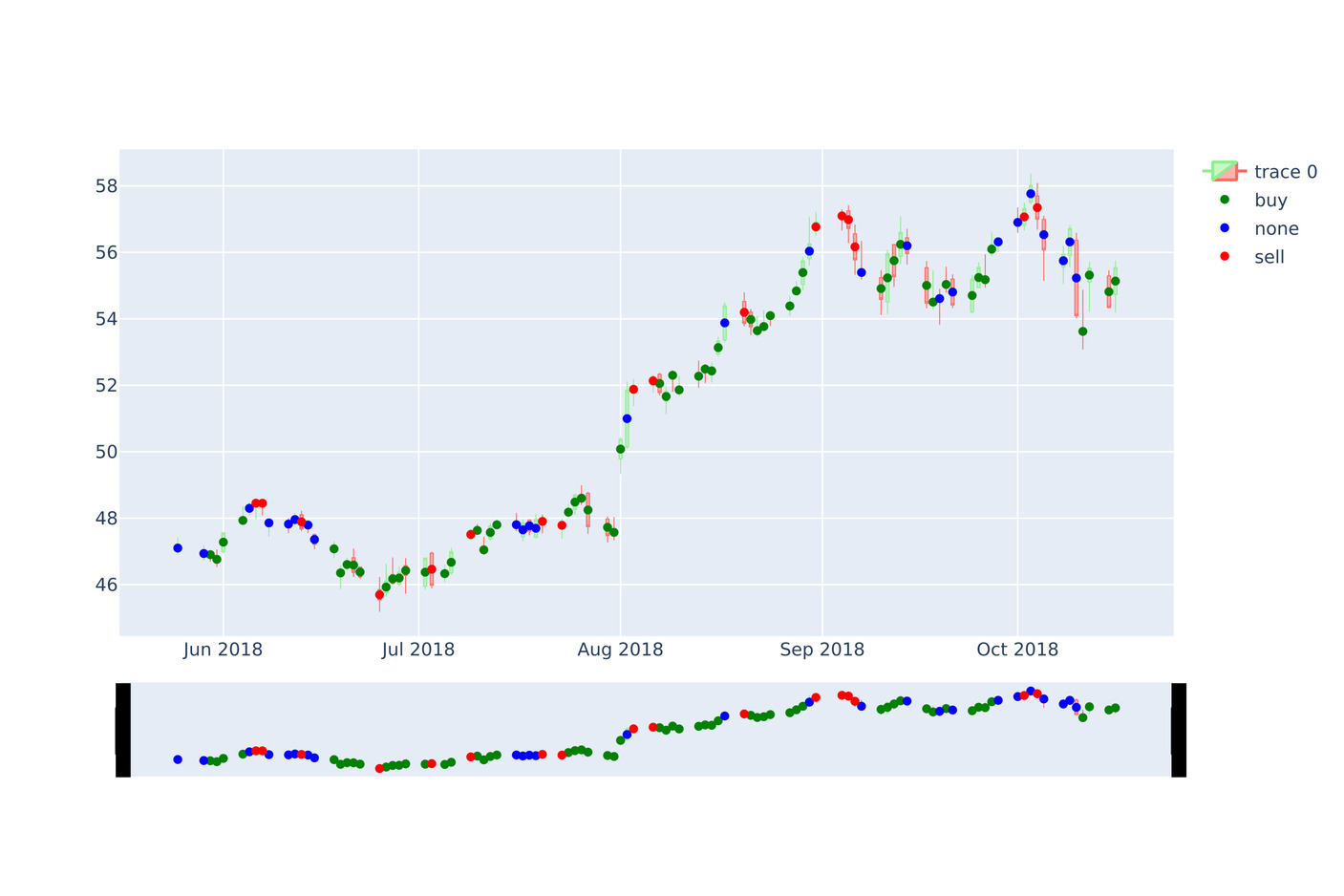}
		\caption{Trading strategy generated for AAPL}
	\end{subfigure}
	\\
	\begin{subfigure}{0.5\textwidth}
		\includegraphics[width=\linewidth, height=0.20\textheight]{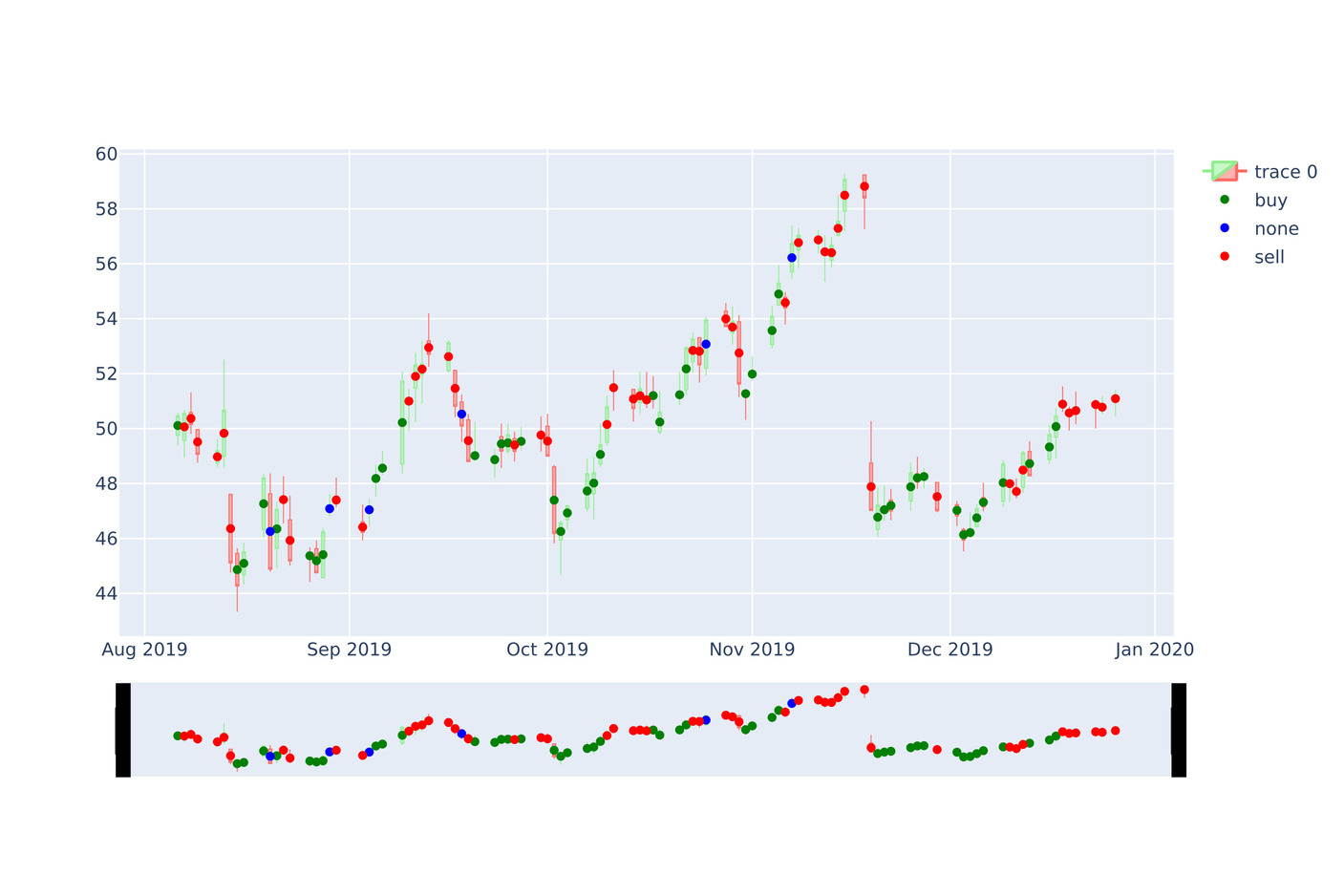}
		\caption{Trading strategy generated for KSS}
	\end{subfigure}%
	\begin{subfigure}{0.5\textwidth}
		\includegraphics[width=\linewidth, height=0.20\textheight]{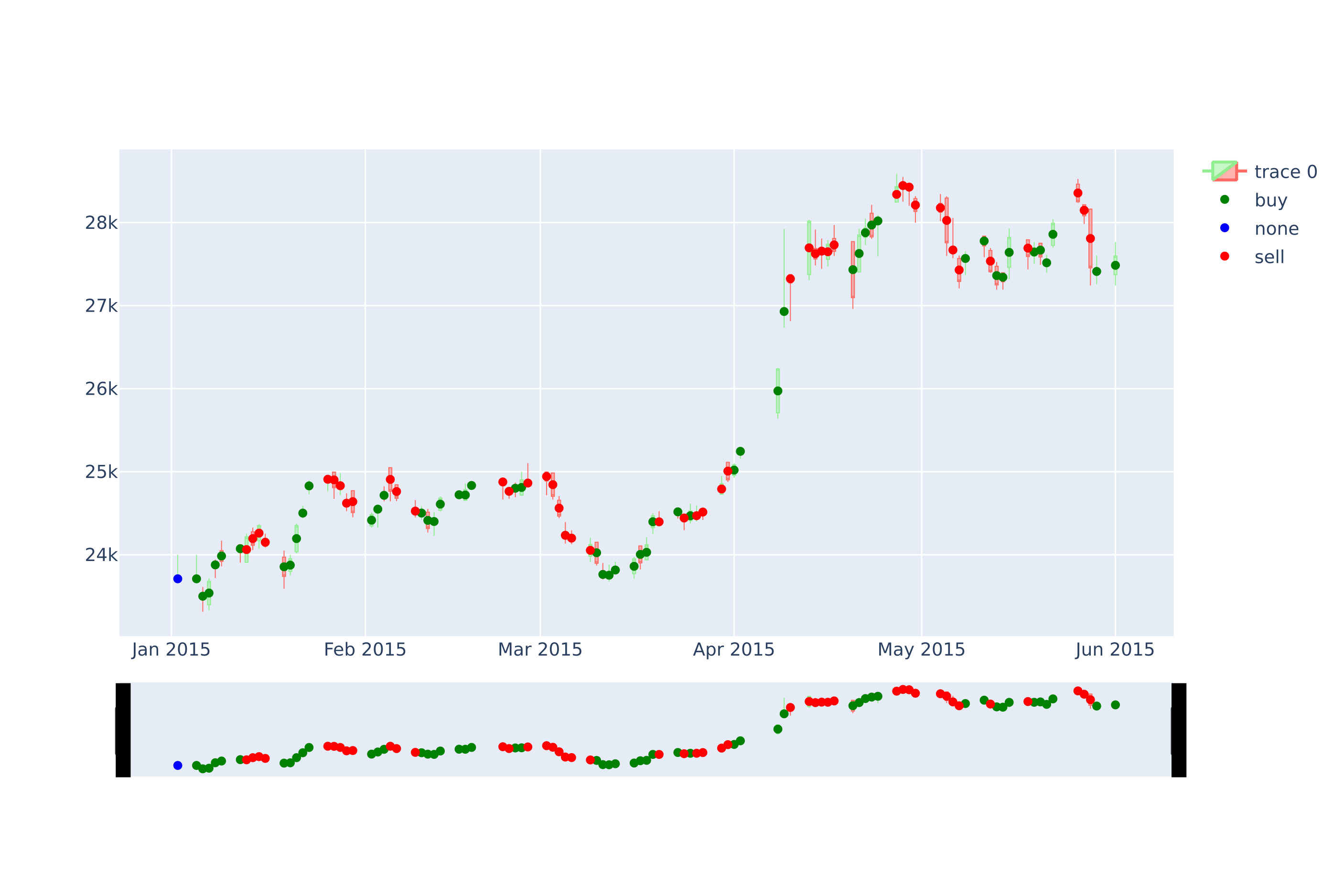}
		\caption{Trading strategy generated for HSI}
	\end{subfigure}
	\\
	\begin{subfigure}{0.5\textwidth}
		\includegraphics[width=\linewidth, height=0.20\textheight]{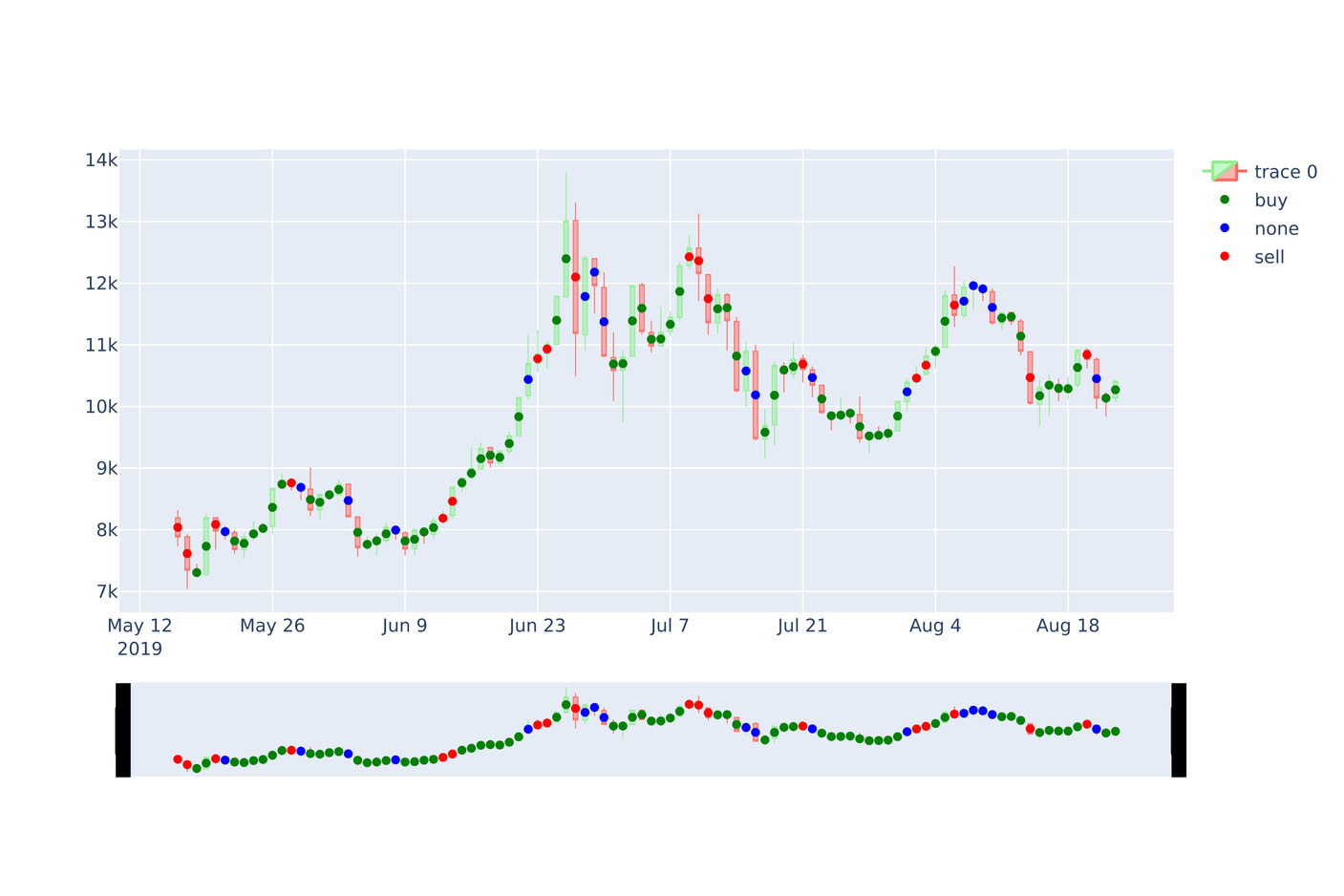}
		\caption{Trading strategy generated for BTC/USD}
	\end{subfigure}
	\\
	\caption{The histogram of strategies generated on each dataset for a period of time with length 100 by the best model.}
	\label{fig:strategies}
\end{figure*}

For each data, the trading strategy is illustrated in \ref{fig:strategies} based on the decisions made at each time step by the most profitable model. The green, red, and blue points represent the 'buy', 'sell', and 'none' signals, respectively. When the agent generates a signal, it will influence the next day's investment. In other words, when the agent decides to buy a share, this action is actually done the next day. As mentioned before, we use a parameter \textit{OwnShare}, which tells us whether the agent already bought the share or not. Thus, when the agent bought a share at the time step $t$, the \textit{OwnShare} parameter would become $true$, and if the next action is \textit{none}, the agent's money will continue to be invested. We begin by an initial investment at $t_0$, and at each time-step, when the agent decides to buy or sell, all the money would be invested or withdrawn. 

As shown in \ref{fig:strategies}, agents could generate signals properly in positions where the trend of the market changes. In order to represent the strategy behavior for each data, we select a period of 100 intervals. The stable markets such as GE, GOOGL, and AAPL contain 'none' signals in their strategy more than volatile markets such as HSI and KSS. That can explain the fact that in stable markets, the market trend changes less rapidly than highly volatile markets; therefore, agents can produce more 'none' action in their strategy.

\subsection{Comparing models with similar works}
Whenever possible, the proposed models in this paper are compared with the state-of-the-art models of learning single asset trading rules. Since most of these models' implementations are not accessible, comparison with each baseline model is accomplished just in cases that the model's performance metrics are reported in the companion paper. The list of the used baseline models is as follows.

\begin{enumerate}[i)]
	\item \textbf{Buy and Hold (B\&H)}:
	
		B\&H is one of the most widely used benchmark strategies to compare the performance of a model. In this strategy, the investor selects an asset and buys it at the first time step of the investment. The purchased asset is held to the end of the period regardless of its price fluctuations.
	\item \textbf{GDQN}:
	
		Proposed by Wu et al. \cite{wu2020adaptive}, uses the concatenation of the technical indicators and raw OHLC price data of the last nine time steps as the input, a two-layered stacked structure of GRUs as the feature extractor, and the DQN as the decision-making module.
		
	\item \textbf{DQT}:
	
		Proposed by Wang et al. \cite{wang2017deep}, implements online Q-learning algorithm to maximize the long-term profit of the investment using the learned rules on a single financial asset. The reward function here is formed by computing the accumulated wealth over the last $n$ days.
	\item \textbf{DDPG}:
	
		Proposed by Xiong et al. \cite{xiong2018practical} uses Deep Deterministic Policy Gradient(DDPG) as the deep reinforcement leaning approach to obtain an adaptive trading strategy. Then, the model's performance is evaluated and compared with the Dow Jones Industrial Average and the traditional min-variance portfolio allocation strategy.
\end{enumerate}

Tables \ref{tbl:wang2017deep-compare}, \ref{tbl:wu2020adaptive-compare}, and \ref{tbl:xiong2018practical-compare} represent our models' performance along with the state-of-the-art models using the profit metrics. According to the results reported in \ref{tbl:wang2017deep-compare}, the performance of the model with MLP encoder and raw OHLC input is significantly better than DQT and RRL on stocks HSI and S\&P500 proposed by Wang et al. \cite{wang2017deep}. For HSI, time-series models achieve a performance close to MLP-vanilla, but they behave poorly on S\&P500.

\ref{tbl:wu2020adaptive-compare} represents the Rate of Return (\%) for our models with different encoders and models proposed by Wu et al. \cite{wu2020adaptive}. Wu et al.'s best model performance is on AAPL stock with Rate of Return equal to 77.7, but the GRU model gains the Rate of Return 438, which is significantly better. Moreover, wherever the models proposed by Wu et al. got a negative return, our model returns a highly positive return. Consider stock GE, where the maximum return value in Wu et al. is -6.39\%, but the best strategy proposed by MLP with vanilla input is 130.4\%. When examining the returns gained by different models on IBM, it is obvious that return values for time-series models are better than those with raw OHLC input, and the GRU encoder gains the highest return of 174\%. This concept explains the existence of a temporal relationship in IBM stock in that specific period. 

\ref{tbl:xiong2018practical-compare} shows the performance of DDPG, the model presented by Xiong et.al. \cite{xiong2018practical}. The final portfolio value of models in our work is better than DDPG, starting with an initial portfolio value of 10000. The CNN-GRU has the best performance with a final portfolio value of 21984, while the DDPG model's final portfolio value is 19791.

As the results indicate, our models perform significantly better than similar models in profitability, ranging from time-series models to raw OHLC inputs. As previously mentioned, these papers' codes were not available, and we had to compare the performance according to common metrics.      

\begin{table}[htb]
	\centering
	\caption{Compare profitability performance with Wang et. al. \cite{wang2017deep} based on Accumulated Return(\%)}
			\begin{tabular}{|c|c|c|}
				\hline
				Agents &‌HSI & S\&P500\\
				\hline
				MLP-vanilla & 13231.2 & 5032.3 \\
				DQN-vanilla & 5016 & 2524\\
				MLP-windowed & 7227 & 4118\\
				DQN-windowed & 7576 & 4289\\
				GRU & 10911 & 3918 \\
				CNN & 10575 &‌ 3859\\
				CNN-GRU & 12566 & 2573\\
				B\&H & 153.5 & 168.6\\
				\hline
				\hline
				B\&H & 154 & 169\\
				\hline
				DQT & 350 & 214\\
				\hline
				RRL & 174 & 141\\
				\hline
			\end{tabular}
\label{tbl:wang2017deep-compare}	
\end{table}

\begin{table}[htb]
	\centering
	\caption{Compare profitability performance with Wu et. al. \cite{wu2020adaptive} based on Rate of Return(\%)}

			\begin{tabular}{|c|c|c|c|c|c|}
				\hline
				Agents & AAPL & GE &‌AXP &‌CSCO & IBM\\
				\hline
				DQN-vanilla & 336 & 129 & 183 & 182 & 144\\
				MLP-vanilla & 262.3 & 130.4 & 260.2 & 259.9 & 149.2\\
				DQN-windowed & 425 & 70 & 252 & 241 & 118\\
				MLP-windowed & 402 & 74 & 280 & 299 & 165\\
				GRU & 438 & 129 & 262 & 233 & 174\\‌
				CNN & 290 & 84 & 189 & 251 & 153\\
				CNN-GRU & 411 & 78 & 284 & 227 & 152\\
				\hline
				\hline
				GDQN & 77.7 & -10.8 & 20.0 & 20.6 & 4.63\\
				\hline
				GDPG & 82.0 & -6.39 & 24.3 & 13.6 & 2.55\\
				\hline
				Turtle & 69.5 & -17.0 & 25.6 & -1.41 & -11.7\\
				\hline
			\end{tabular}
	\label{tbl:wu2020adaptive-compare}
\end{table}

\begin{table}[htb]
	\centering
	\caption{Compare profitability performance with Xiong et. al. \cite{xiong2018practical} based on Final Portfolio Value(Initial Portfolio Value is 10000)}
			\begin{tabular}{|c|c|}
				\hline
				Agents & DJI\\
				\hline
				DQN-vanilla & 20275 \\
				MLP-vanilla & 21580\\
				DQN-windowed & 19475 \\
				MLP-windowed & 20760\\
				GRU & 21360\\
				CNN & 20287\\
				CNN-GRU & 21984 \\
				\hline
				\hline
				DDPG & 19791\\
				\hline
				Min-Variance & 14369\\
				\hline
				DJIA & 15428\\
				\hline
			\end{tabular}
			\label{tbl:xiong2018practical-compare}
\end{table}

\section{Conclusion}
In this work, we proposed a method based on the Encoder-Decoder framework, where the encoder model is a DNN, which helps extract essential features from the raw financial data, and the decoder is a DRL agent which makes a decision at each time-step and generates trading signals. The model is trained end-to-end, and the encoder's feature extraction function is optimized toward the policy improvement of the DRL agent. 

The DRL is based on the Deep Q-learning algorithm and consists of a policy and a target network, both of which are multi-layered Perceptron networks. For the encoder part, the feature extraction performance of various DNNs is evaluated and compared. The proposed models for the encoder part are categorized into two types: 
1) The raw OHLC input, which receives candle OHLC prices directly. 
2) Time-series input, which concatenates a window of consecutive candles and receives the window as input.

Based on experimental results, the performance of models depended on the market behavior. When the market is highly volatile, meaning that the rate of price fluctuation is high, DQN and MLP model with the raw OHLC input had the best performance since they are able to make decisions only based on current input representation, disregarding to the historical changes of the market. On the other hand, there are more stable markets where models with time-series input can devise more profitable trading strategies because the market behavior enables them to exploit efficient features from financial data history.
The window size impact is further studied, and we concluded that window sizes in the interval of 10 to 20 have the best feature extraction performance.
Using the trading strategies generated for each data, we understand that agents working on stable stocks will generate \textit{none} signal more frequently than in strategies devised for highly volatile markets.

The future of the work can be viewed from different perspectives. 
\begin{itemize}
	\item As we have experimented, if we could predict the next state of the environment using the current state, and feed the predicted next state to the DRL model, the performance would significantly increase.
	\item The actor and actor-critic based DRL methods can be tested and compared with the performance of critic based Deep Q-learning algorithm used here.
	\item Working on offering a metric used to describe the behavior of the market, based on which, we can specify whether the time-series models can work efficiently in rule extraction or not. Using this metric, we can distinguish where to apply models with raw OHLC input or apply time-series input.
\end{itemize}

\section*{References}

\bibliography{references}

\end{document}